# Radio frequency emissions from dark-matter-candidate magnetized quark nuggets interacting with matter

J. Pace VanDevender[1*], C. Jerald Buchenauer[2], Chunpei Cai[3], Aaron P. VanDevender[4], and Benjamin A. Ulmen[5]

[1]VanDevender Enterprises LLC, 7604 Lamplighter LN NE, Albuquerque, NM 87109 USA.
[2]Department of Electrical and Computer Engineering, MSC01 1100, University of New Mexico, Albuquerque, NM 87131 USA. [3]Department of Mechanical Engineering and Engineering Mechanics, Michigan Technological University, MEEM 1013, 1400 Townsend Drive, Houghton, Michigan 49931, USA. [4]Founders Fund, One Letterman Drive, Building D, 5th Floor, Presidio of San Francisco, San Francisco, CA 94129 USA. [5]PO Box 5800, MS-1159, Sandia National Laboratories, Albuquerque, NM 87115-1159

*pace@vandevender.com

**Abstract**

**Quark nuggets are theoretical objects composed of approximately equal numbers of up, down, and strange quarks. They are also called strangelets, nuclearites, AQNs, slets, Macros, and MQNs. Quark nuggets are a candidate for dark matter, which has been a mystery for decades despite constituting ~85% of the universe's mass. Most previous models of quark nuggets have assumed no intrinsic magnetic field; however, Tatsumi found that quark nuggets may exist in magnetars as a ferromagnetic liquid with a magnetic field $B_S = 10^{12\pm1}$ T. We apply that result to quark nuggets, a dark-matter candidate consistent with the Standard Model, and report results of analytic calculations and simulations that show they spin up and emit electromagnetic radiation at ~$10^4$ to ~$10^9$ Hz after passage through planetary environments. The results depend strongly on the value of $B_o$, which is a parameter to guide and interpret observations. A proposed sensor system with three satellites at 51,000 km altitude illustrates the feasibility of using radio-frequency emissions to detect MQNs during a five year mission.**

## Introduction

About 85% [1] of the universe's mass does not interact strongly with light; it is called dark matter, is distributed in a halo throughout a galaxy [2]. Identifying the nature of dark matter is currently one of the biggest challenges in science. As reviewed most recently by Salucci [3], extensive searches for a subatomic particle that would be consistent with dark matter have yet to detect anything above background signals.

Most models assume dark matter interacts with normal matter only through gravity and the weak interaction. However, detailed analysis of the accumulating data on galaxies of different types suggest dark matter and normal luminous matter interact somewhat more strongly, on the time scale of the age of the Universe [3]. Magnetized quark nuggets (MQNs) [4] are an emerging candidate for dark matter that quantitatively meets the traditional interaction requirements for dark matter and still interacts with normal matter on the time scale of the age of the Universe through the magnetic force. In this paper we investigate a new method for detecting MQNs and measuring their properties.

Quarks are the basic building blocks of protons, neutrons, and many other particles in the Standard Model of Particle Physics [5]. Macroscopic quark nuggets [6], which are also called strangelets [7], nuclearites [8], AQNs [9], slets [10], and Macros [11] are theoretically predicted objects composed of up, down, and strange quarks in essentially equal numbers. A brief summary of quark-nugget research [6-35] on charge-to-mass ratio, formation, stability, and detection has been updated from Ref. 16 and is provided for convenience as Supplementary Note: Quark-nugget research summary.

Most previous models of quark nuggets have assumed negligible self-magnetic field. However, Tatsumi [15] explored the internal state of quark-nugget cores in magnetars and found that quark nuggets may exist as a ferromagnetic liquid with a surface magnetic field $B_S = 10^{12\pm1}$ T. Although his calculations used the MIT bag model with its well-known limitations [20], his conclusions can and should be tested. We have applied his ferromagnetic-fluid model of quark nuggets in magnetars to magnetized quark nuggets (MQNs) [4,16] dark matter and extend those results to calculate how MQNs rotate and radiate radio frequency (RF) emissions during and after interaction with normal matter. We also explore how the RF emissions can enable detection of MQNs.

As done in references 24 and 44, we will use $B_o$ as a key parameter. The value of $B_o$ is related to the mean value $<B_S>$ of the surface magnetic field through

$$<B_S> = \left(\frac{\rho_{QN}}{10^{18}(kg/m^3)}\right)\left(\frac{\rho_{DM\_T=100Mev}}{1.6 \times 10^8(kg/m^3)}\right)B_o. \qquad (1)$$

If MQN mass density $\rho_{QN} = 10^{18}$ kg/m³ and the density of dark matter $\rho_{DM} = 1.6 \times 10^8$ kg/m³ at time t ≈ 65 μs (when the temperature T ≈ 100 MeV in accord with standard ΛCDM cosmology), then $B_o = <B_S>$. If better values of $\rho_{QN}$, and $\rho_{DM}$ are found, then the corresponding values of $<B_S>$ can be calculated with equation (1) from those better values and from $B_o$ determined by observations.

Previous papers on MQNs showed:

1) self-magnetic field aggregates MQNs with baryon number $A = 1$ into MQNs with a broad mass distribution [4] that is characterized by the value of $B_o$ and typically has $A$ between ~$10^3$ and $10^{37}$,

2) aggregation dominates decay by weak interaction so massive MQNs can form and remain magnetically stabilized in the early universe even though they have not been observed in particle accelerators [4],

3) the self-magnetic field forms a magnetopause that strongly enhances the interaction cross section of a MQN with a surrounding plasma [16] that is primarily sustained by radiation and electron impact ionization in the high temperature plasma formed by normal matter stagnating against the magnetopause,

4) $B_o > 3 \times 10^{12}$ T is excluded [4] by the lack of observed deeply penetrating impacts that deposit > Megaton-TNT equivalent energy per km, so Tatsumi's range of $B_S$ is reduced to $1 \times 10^{11}$ T $\leq B_o \leq 3 \times 10^{12}$ T,

5) MQNs satisfy criteria for dark matter even with baryon's interacting with MQNs through the self-magnetic field [4], and

6) the unexcluded range of $B_o$, the low (~$7 \times 10^{-22}$ kg m$^{-3}$) density of local dark matter, net incident velocity of ~250 km/s, and the high average mass of MQNs constrain the flux of MQNs of all masses to between $10^{-7}$ and $4 \times 10^{-15}$ m$^{-2}$ y$^{-1}$ sr$^{-1}$ and constrain the flux for MQN masses >1 kg to between $3 \times 10^{-14}$ and $2 \times 10^{-17}$ m$^{-2}$ y$^{-1}$ sr$^{-1}$, so very large area detectors or very long observation times are required [4] to detect MQNs.

In this paper, we investigate the possibility of detecting MQNs interacting with the largest accessible target: Earth, with its magnetosphere to 10 Earth radii. We show MQNs necessarily experience a net torque while passing through matter, can spin up to kHz to GHz frequencies, and emit sufficient narrow-band electromagnetic radiation that could be detected in additional tests of the MQN hypothesis for dark matter.

Detection is also complicated by the magnetopause plasma (hot ionized gas) shielding some radio-frequency (RF) emissions. Spacecraft re-entering the atmosphere cannot communicate with ground stations because the surrounding plasma shields the emissions until the spacecraft slows down. The same blackout effect prevents RF emissions from MQNs from being detected during the high-velocity interaction of MQMs with matter in the troposphere or ionosphere.

However, MQNs that exit the atmosphere can be detected by their narrow-band, time-varying RF radiation, with frequency equal to the rotation frequency. In addition, MQN detection rate should be strongly correlated with the direction of dark-matter flux into the detector aperture. That preferred direction is determined by Earth's velocity about the galactic center and, consequently, through the dark-matter halo. These characteristics should permit their detection and differentiation from background RF.

**Results**

In this section, we show 1) when MQNs interact with plasma, their rotational velocity increases as their translational velocity decreases, 2) MQNs passing through Earth's atmosphere on a fly-by trajectory produce sufficient RF emissions to be detected by satellites but not ground-based sensors, and 3) the frequency, RF power, and event rate depend strongly on the value of the $B_o$ parameter in equation (1).

**Rotational spin-up**

Plasma is an ionized state of matter with sufficient electron number density $n_e$ at temperature $T_e$ for the electrons and ions to behave as a quasi-neutral fluid. Quantitatively, that means there is at least one electron-ion pair in a spherical volume of radius equal to a Debye shielding length $\lambda_D$,

$$\lambda_D = \sqrt{\frac{\varepsilon_o k_B T_e}{n_e e^2}}, \tag{2}$$

in which the permittivity of free space $\varepsilon_o = 8.854 \times 10^{-12}$ F/m, the Boltzmann constant $k_B = 1.38 \times 10^{-23}$ J/K, and electron charge $e = 1.6 \times 10^{-19}$ C. The Debye length is also the distance over which the thermal energy of the electrons permits charge separation to occur and defines electrical quasi-neutrality of a plasma [36] versus an ensemble of electrons and ions which behave as particles. The distinction is important for MQNs. A plasma impacting the magnetic field of a MQN acts as a conducting fluid and compresses the magnetic field to form a magnetopause [16]. A co-moving electron-ion pair impacting the magnetic field will separate because the forces from their opposite charges deflect them in opposite directions and will create a local electric field between them. The combined force from their electric $\boldsymbol{E}$ and magnetic $\boldsymbol{B}$ fields causes them to move in the same $\boldsymbol{E}\times\boldsymbol{B}$ direction at the same velocity [36] and through the magnetic field, instead of strongly compressing it. In contrast, conducting plasma shorts out the electric field and prevents particle penetration.

A MQN moving through a plasma experiences a greatly-enhanced slowing down force [16] through its magnetopause, which is the magnetic structure formed by particle pressure from a plasma stream balancing magnetic field pressure around a magnetic dipole. For example, the solar wind forms a magnetopause with Earth's magnetic field. Since the particles' mean free path for collisions is much larger than the Larmor radius in the magnetic field, the physics of Earth's magnetopause is collisionless and applicable to the very small-scale lengths of a quark-nugget's magnetopause.

As derived in Ref. 16, the cross section $\sigma_m$ for momentum transfer by the magnetopause effect is

$$\sigma_m = \pi r_m^2 = \pi \left( \frac{2 B_o^2 r_{QN}^6}{\mu_0 \rho_x v^2} \right)^{\frac{1}{3}} \tag{3}$$

for magnetopause radius $r_m$, MQN radius $r_{QN}$, MQN speed $v$, and mass density of surrounding matter $\rho_x$. The total force $F_o$ exerted by the plasma on the quark nugget is approximately

$$F_o \approx \sigma_m \rho_x v^2. \tag{4}$$

Equations (3) and (4) let us calculate the decelerating force on the MQN and its translational velocity during passage through matter [16].

In this paper, we extend the dynamics to calculate the torque and rotational velocity of the MQN passing through matter. Papagiannis [37] showed that the solar wind, which has mass density $\rho_x \approx 10^{-20}$ kg/m$^3$ and velocity $v \approx 3.5 \times 10^5$ m/s, exerts a torque $T$ (N · m) on Earth (radius $r_o = 6.37 \times 10^6$ m and magnetic field $B_o = 3. \times 10^{-5}$ T) as a function of the angle $\chi$ between the magnetic axis and the normal to both the magnetic axis and the direction of the solar wind. His semi-empirical result is expressed in MKS units as

$$T = C_2 \rho_x^{0.5} v B_o r_{QN}^3 \tan(\chi), \tag{5}$$

in which $C_2 = 1400$ with units of $Ns(kg)^{-0.5} m^{-1.5} T^{-1}$. Papagiannis validated the expression for the angles $\chi$ within 0.61 radians of 0 and within 0.61 radians of $\pi$ (i.e. $-0.61 \leq \chi \leq +0.61$, $-3.14 \leq \chi \leq -2.53$, and $+2.53 \leq \chi \leq +3.14$, as illustrated in Fig. 1). By symmetry of the magnetic field, the torque is 0 at $\chi = 0$ and $\chi = \pm \pi/2$.

Since Papagiannis was only considering Earth, he limited his calculations to within 0.61 radians of the normal to the plasma velocity. Rigorously reproducing and extending his computational results to the larger angles required for MQN rotation is beyond the scope of this paper. Therefore, we extend his results by observing the amplitude of the torque is symmetric about $\chi = -3\pi/4, -\pi/4, \pi/4$ and $3\pi/4$, as shown in Fig. 1 by solid blue lines, and approximate the rest of the torque function by extrapolation of the $\tan(\chi)$ function in equation (5), as shown with dotted blue lines in Fig. 1. The resulting functions $F_\chi$ and T, shown in equation (6), replace equation (5).

$$\begin{aligned}F_\chi &= MIN(ABS(\tan \chi), ABS(\cot \chi)) \frac{\tan \chi}{ABS(\tan \chi)} \\ T &= C_2 \rho_x^{0.5} v B_o r_{QN}^3 F_\chi\end{aligned} \tag{6}$$

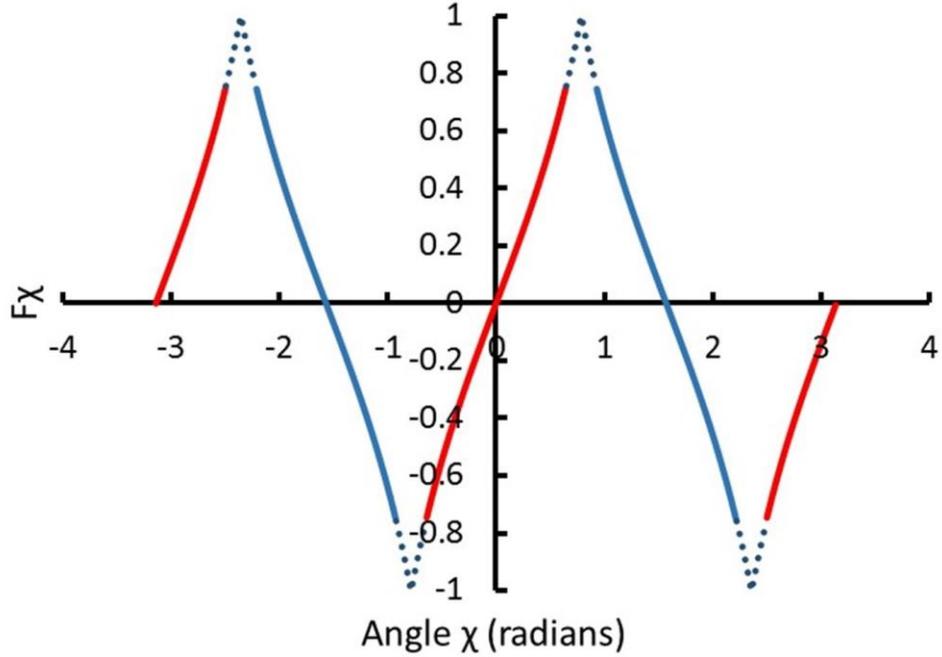

**Figure 1. Generalization of the tan(χ) factor in equation (5), in which χ is the angle between the magnetic axis and the normal to both the magnetic axis and the quark-nugget's direction of travel in the rest frame of the quark nugget.** The solid red lines indicate the angles computed by Papagiannis; the solid blue lines indicate extensions by symmetry. Dotted blue lines indicate functional extrapolation of Papagiannis and symmetry-extension values.

The torque is negligible for Earth but is very large for a quark nugget. The rate of change of angular velocity $\omega$ for MQN with mass $m_{QN}$, moment of inertia $I_{mom} = 0.4\, m_{QN}\, r_{QN}^2$ experiencing torque $T$ is

$$\frac{d\omega}{dt} = \frac{T}{I_{mom}} \quad . \tag{7}$$

Equations (2) through (7) were solved for the angular velocity versus time. The interaction produces a velocity-dependent and angle-dependent torque that causes MQNs to oscillate initially about an equilibrium. Since the quark nugget slows down as it passes through ionized matter, the decreasing forward velocity decreases the torque with time, so the time-averaged torque in one half-cycle is greater than the opposing time-averaged torque in the next half-cycle. The amplitude of the oscillation necessarily grows, as shown in Fig. 2. Once the angular momentum is sufficient to give continuous rotation, the net torque continually accelerates the angular motion to produce a rapidly-rotating quark nugget. As shown in Fig. 2, MHz frequencies are quickly achieved even with a 0.1 kg quark nugget moving through 1 kg/m³ density air at 250 km/s. For smaller or larger masses, the resulting angular acceleration and velocity are respectively larger or smaller.

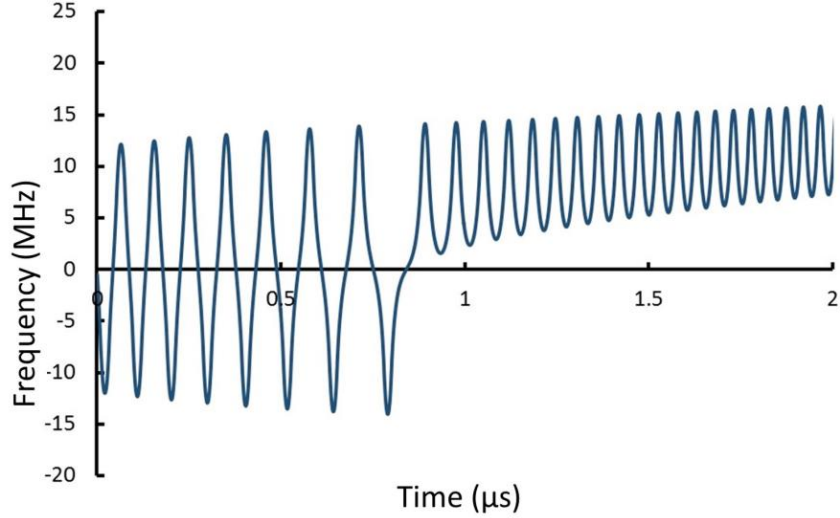

**Figure 2. Calculated frequency versus time for 0.1 kg MQN with $B_o = 2.25 \times 10^{12}$ T, initial velocity of 250 km/s, and passing through air at density 1.0 kg/m³.** Note the initial oscillation is about 0 until angular momentum becomes sufficient to complete a full rotation.

Several other approximations for the torque in the intervals shown with dotted lines in Fig. 1 gave the same frequency within 5%., which is within the uncertainty of the magnetic field parameter $B_o$.

**Equilibrium frequency and radiated power**

Rotating magnetic dipoles emit electromagnetic radiation in the far field with power per steradian [38] given by

$$\frac{dP}{d\Omega} = \frac{Z_o}{32\pi^2}\left(\frac{\omega}{c}\right)^4 m_m^2 \sin^2 \chi \tag{8}$$

in SI units, with $Z_o = 377$ Ω, $\omega$ is angular frequency, and $c$ is the speed of light in vacuum. The magnetic dipole moment $m_m = 4\pi B_o r_{QN}^3/\mu_o$, and angle of rotation $\chi$ is the angle between the velocity of the incoming plasma and the magnetic moment. The total power radiated [38] is

$$P = \frac{Z_o}{12\pi}\left(\frac{\omega}{c}\right)^4 m_m^2. \tag{9}$$

The spin-up process strongly depends on the details of the surrounding material mass and MQN velocity along the path of the MQN, the MQN mass, and surface magnetic field $B_o$. In spite of these complexities, we find that the spin-up time is very much less than the MQN transit time through the region of highest torque, and the lower limit of final rotation frequency can be adequately estimated by assuming the energy gained per cycle equals the energy radiated per cycle:

$$\int_0^{\frac{2\pi}{\omega}} T dt = \frac{2\pi P}{\omega^2}, \tag{10}$$

in which the torque $T$ is given by equation (6) and the radiated power $P$ is given by equation (9). Combining equations (6) through (10) gives

$$\int_0^{\frac{2\pi}{\omega}} \frac{v(t) F_\chi(x(\chi(\omega t)))}{\omega^2} dt = \frac{5.54 \times 10^{-22} B_o r_{QN}^3}{\rho_x^{0.5}}. \tag{11}$$

In the simulations discussed below, we solve equations (3) and (4) for position and velocity $v(t)$ calculated along a trajectory through the atmosphere to the position of maximum density $\rho_x$, where we solve equation (11) for the lower-limit to the maximum frequency $\omega_{max}$.

**Attenuation of RF power by magnetopause plasma**

The surrounding magnetopause plasma has a characteristic plasma frequency

$$\omega_{pe} = \left(\frac{4\pi n_e e^2}{m_e}\right)^{\frac{1}{2}} = 56.4 n_e^{\frac{1}{2}}, \tag{12}$$

in which $n_e$ = the local electron number density, $e$ = the electron charge, and $m_e$ = the electron mass. The characteristic e-fold length for attenuating the radiated power for frequency $\omega < \omega_{pe}$ is $k_{atten}^{-1}$:

$$k_{atten}^{-1} = \frac{c}{2\omega_{pe}} \frac{1}{\sqrt{1 - \left(\frac{\omega}{\omega_{pe}}\right)^2}}. \tag{13}$$

In practice, the scale length $0.5c/\omega_{pe}$ is approximately 0.5% of the ion Larmor radius, which is approximately the minimum thickness of the magnetopause boundary. Therefore, magnetopause plasma strongly absorbs RF energy for frequencies less than the plasma frequency.

Since plasmas strongly absorb electromagnetic radiation with frequency less than the plasma frequency, which varies with solar activity, practical detection of MQNs by their RF emissions is limited to ≥0.03 MHz in the solar-wind plasma near Earth orbit, to ≥0.4 MHz in the magnetosphere, and to ≥40 MHz in the ionosphere. The equilibrium frequency of the MQN spin-up process and the high density of the troposphere means, in practice, all RF emissions in the troposphere are strongly shielded. Therefore, we will focus on detecting MQNs in the magnetosphere after they have transited Earth's atmosphere, as illustrated in Fig. 3.

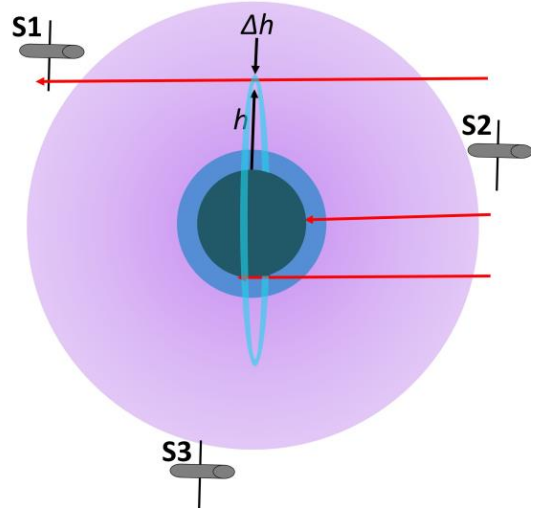

**Figure 3. Near Earth environment with MQN trajectories (red lines).** Three concentric circles represent MQN interaction volume: highly ionized and low density magnetosphere and ionosphere (purple); weakly ionized or neutral, low-density troposphere (blue), and neutral, high-density planet (gray). Satellites S1, S2 and S3 are shown in orbits that let them monitor narrow-band RF emissions above background as described below. The vertical ring (light blue) is a detection-area element for simulations of MQNs interacting with magnetosphere, ionosphere and troposphere, as discussed below.

The middle MQN trajectory in Fig. 3 represents a direct impact [16]. The bottom trajectory represents a MQN that is gravitationally captured and does not exit into the magnetopause. The top trajectory represents a MQN that spins up during transit and is detected by satellite S1. Satellite S2 would not detect these MQNs since they have not passed through sufficient matter to spin up. Therefore, appropriate differences in event rates as a function of satellite position would support detection of dark matter.

Trajectories of quark nuggets are from a preferred direction in Fig. 3 because the velocity of the solar system about the galactic center and through the halo of dark matter nearly dominates the random velocity of quark nuggets, as discussed in a subsequent section.

**Radiofrequency background from MQNs distributed throughout the galaxy**

Dark-matter is distributed throughout the galaxy [2,3]. Therefore, MQN dark matter in interstellar space might interact with the local plasma density and emit RF radiation that fills the universe over billions of years. The resulting RF might be detectable as a galactic background near Earth. Olbers' Paradox on why the night sky is dark [39] addresses the same phenomenon for photons from stars. Evaluating the potential for detecting and interpreting this radiation is complex. A detailed analysis is beyond the scope of this paper, which focuses on detection of near-Earth MQNs. However, the following preliminary analysis shows that the MQN hypothesis cannot be tested by observations of galactic RF background near Earth.

The plasma density and temperature in interstellar space vary greatly:

1) a minimum of ~$10^{-4}$ to ~$10^{-2}$ particles/cm$^3$ in Hot Ionized Media (HIM) at $T_e$ ~ $10^6$ to $10^7$ K composing 20% to 70% of interstellar space,

2) to ~0.2 to 0.5 particles/cm$^3$ in Warm Ionized Media (WIM) at $T_e$ ~ 8,000 K composing 20% to 50% of space,

3) to higher mass densities represented by ~$10^2$ to $10^6$ particles/cm$^3$ in molecular clouds at $T_e$ ~ 10 to 20 K composing < 1% of space [40].

As discussed above, matter has to act like a fluid plasma (instead of an ensemble of isolated charged particles) to form a magnetopause. Quantitatively, the scale length $\lambda_D$ for charge separation in equation (2) has to be less than the magnetopause radius $r_m$ defined in equation (3), so

$$\frac{r_m}{\lambda_D} = 4000 \cdot \left(\frac{B_o m_{QN} n_e}{\rho_{QN} v}\right)^{\frac{1}{3}} \left(\frac{1}{T_e}\right)^{\frac{1}{2}} > 1 \ . \tag{14}$$

For example, using the mid-range value of $B_o = 2.0 \times 10^{12}$ T and $\rho_{QN} = 1 \times 10^{18}$ kg/m$^3$, the MQN mass has to be greater than $10^7$ kg for a magnetopause to form in HIM. The equilibrium rotation frequency for such massive MQNs is much less than 30 MHz and the corresponding RF is absorbed by the solar-wind plasma near Earth.

The corresponding threshold MQN mass for forming a magnetopause in WIM is 10 kg. However, the higher density plasma in the WIM slows those MQNs, so their RF is still less than 30 kHz, so it is also absorbed in the solar-wind plasma near Earth.

In general, the larger mass density of the rest of interstellar media produce even lower frequency RF, which is absorbed even further from Earth. We analyzed these effects for the full range of interstellar plasma conditions [40] and MQN mass distributions [4] as a function of the $B_o$ parameter. We found wherever MQN dark matter can produce RF emissions, the emissions are either absorbed in the surrounding plasma or in the solar-wind plasma near Earth. The RF is not detectable near Earth and cannot be used to test the MQN hypothesis. Therefore, we focus on detecting MQNs passing very near Earth and radiating well above the 30-kHz cutoff frequency of the solar-wind plasma.

**Discriminating MQN events from background**

Theoretical profiles of dark matter halos are guided by astrophysical observations, which are consistent with a mass density of dark matter near Earth of about $7 \times 10^{-22}$ kg/m$^3$ ± 70% [14]. This extremely low mass density and the very broad mass distributions [4] mean that the flux of MQNs is very low. Detecting them over the full range of not-excluded values of $B_o$ requires interaction volumes of at least planetary size and requires a means of reliably subtracting background. The orientation of the sensed MQN flux with respect to the direction of inflowing dark matter provides one such opportunity.

Models differ in their degree of self-interaction. Computer simulations [41] covering a substantial portion of these models indicate dark matter occupies a halo within and around the galaxy's ordinary matter and has a Maxwellian-like, isotropic velocity distribution. Although simulations predict the velocity distribution of dark matter varies somewhat with the self-interaction model, the most probable, isotropic speed is ~220 km/s with a full-width-at-half-maximum of ~275 km/s.

The solar system moves through this high-speed dark-matter halo in its ~250 km/s motion about the galactic center. The direction of this motion is towards the star Vega, which has celestial coordinates: right ascension 18h 36m 56.33635s, declination +38° 47′ 01.2802″. In addition, Earth moves around the Sun at ~30 km/s. The vector sum of these two velocities gives the net velocity of Earth through dark matter, and the negative of this vector sum is the velocity of dark matter relative to Earth, shown in Fig. 4 for the position of Earth on the first day of each month.

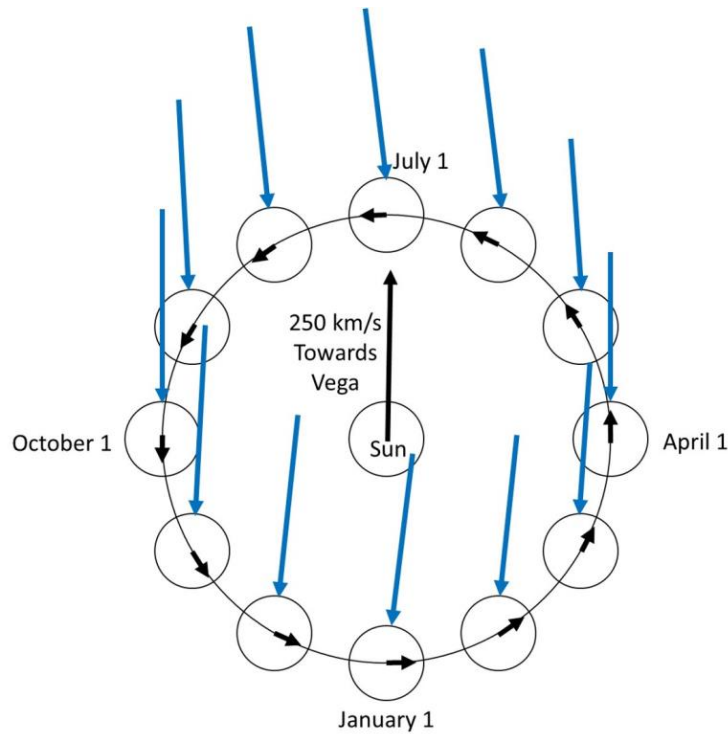

**Figure 4. For the first day of each month, Earth's position and velocity about the Sun are shown.** The solar system's velocity towards Vega is shown by the black vector from the Sun. The net velocity vector of dark matter into Earth is shown in blue for each month. The effects of the 23.5° angle between Earth's equatorial plane and the ecliptic and the 38.8° angle between Earth's equatorial plane and Vega's position are not shown.

A sensor on Earth should detect the most events per hour when it is sensitive to the flux of dark matter from the direction of Vega and much less when Earth shields the detector from the flux. A satellite in orbit about Earth would encounter a higher flux of MQNs when it is not shielded by Earth and when it is within range of MQNs that have transited through enough matter to spin up, as illustrated by S1 in Fig. 3.

The dark-matter velocity distribution has a streaming component $U_s$ relative to the sensor and an isotropic component $U_{iso}$ relative to the galactic center. The isotropic component smooths the transition between the directly exposed and Earth-shielded conditions. $U_{iso}$ can be adequately approximated for our purposes as the highest probability thermal speed $U_{iso} = \sqrt{\dfrac{2kT}{M}}$ of a 3D Maxwellian velocity distribution of dark matter with mass $M$ and with temperature $T$, where $k$ is the Boltzman constant. Simulations [41] indicate that $U_s \approx U_{iso}$. Therefore, we calculated the detection rate as a function of the sensor's orientation on Earth with respect to Vega and $S = U_s/U_{iso}$ using the method developed by Cai, et al. [42]. The results are shown in Fig. 5.

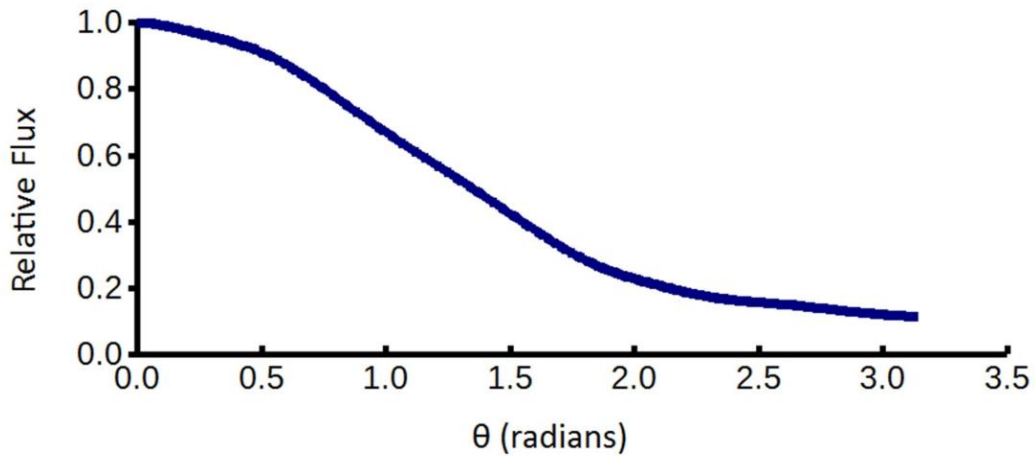

**Figure 5. Calculated and normalized variation of dark matter flux streaming from the direction of the star Vega as a function of polar angle $\theta$ from Vega's zenith.**

If this variation with respect to Vega's position is observed, *i.e.* the event rate for satellite S2 in Fig. 3 is appropriately and systematically less than the event rate for S1, the result would be convincing evidence of having detected MQN dark matter.

**Including MQNs from all directions**

The flux in Fig. 5 is normalized to $F_{\theta=0}$, the total number of events m$^{-2}$ y$^{-1}$ sr$^{-1}$ for a detection surface facing directly into the streaming velocity $U_s$, and was approximated from the mass distributions in Ref. 4, assuming the mean incoming velocity $U_s = 2.5 \times 10^5$ m/s and assuming the effect of $U_{iso} \neq 0$ on the flux is negligible to first order for $\theta = 0$. Assuming cylindrical symmetry in azimuthal direction $\varphi$ and integrating the curve in Fig. 5 over solid angle with $d\Omega = \sin\theta \, d\theta \, d\varphi$ gives the correction factor for estimating the event rate in m$^{-2}$ y$^{-1}$ for MQNs incident from all directions

$$F_{all\_\Omega} = 2\pi \int_0^\pi F_{\theta=0} d\theta = 5.56 \ . \tag{15}$$

The results of the simulation for $\theta = 0$ in subsequent sections are multiplied by 5.56 to estimate the event rate for MQNs from all directions.

**Simulating MQNs flying by Earth**

Consider the simple case of a quark-nugget with a trajectory parallel to a tangent to Earth, as illustrated by the three trajectories in Fig. 3. Satellites sense MQNs after they have transited the highest density matter along their trajectories and experienced the corresponding torque, as described in equation (6), to produce the maximum frequency and radiated power. After they pass into the magnetosphere, they can be detected since their emissions above ~ 0.4 MHz are no longer shielded by the higher-density plasma of the ionosphere.

In our simulations, Earth's atmosphere is divided into increments $\Delta h$ of altitude $h$. MQN trajectories are characterized by their minimum altitude for $0 < h \leq 9 \, r_e$, for $r_e = 6.378 \times 10^6$ m, which is one Earth radius. For each increment $\Delta h$, test MQNs with masses consistent with the mass distributions of Ref. 4 are injected from the right in Fig. 3 with initial velocity in the $x$ direction $v_x = -U_s = -2.5 \times 10^5$ m/s. Their positions and velocities are calculated under the combined effects of gravity and magnetopause interaction. For each test particle, the maximum torque encountered in its trajectory, *i.e.* the torque in equation (6) at the maximum value of the product of total velocity and the square root of the local mass density, is used to calculate the equilibrium frequency from equation (11) and the corresponding RF power from equation (9).

Characteristic times $\tau_{up} = \omega_{max} I_{mom}/T_{max}$ for spin up and $\tau_{down} = 0.5 \, I_{mom} \, \omega_{max}^2/P_{max}$ for spin down by radiation loss are calculated. Since we find $\tau_{up}$ is much less than transit time through the atmosphere, $\omega_{max}$ is the lower limit to the maximum frequency, from equation (11). The value of $\tau_{down}$ helps determine the detectability of each representative MQN. Characteristic e-fold times for spin up and spin down are included in Supplementary Results: Representative Data Tables for Sensor Design. Values of $\tau_{down}$ vary from a minimum of $2.4 \times 10^3$ s to a maximum of $1.9 \times 10^6$ s and provide adequate time for detection.

The cylindrically symmetric cross sectional area $A_h$ associated with the altitude increment $\Delta h$ and altitude $h$ above Earth radius $r_e$ is illustrated in Fig. 3 and is given by

$$A_h = 2\pi(r_e + h)(\Delta h). \tag{16}$$

MQNs with final velocity exceeding low-Earth orbital velocity ≥7400 m/s escape the RF-absorbing ionosphere and will be recorded by a satellite-based sensor.

Atmospheric density as a function of altitude $h = r - r_e$ for MQNs at radius $r$ and Earth radius $r_e$ were derived from the literature and fit with the following equations:

For radius $r$ below the magnetosphere [43], i.e. $0 > r - r_e > 2.873 \times 10^5$ m:

$$\rho_{atm} = 1.0\exp(-(r-r_e)/7.25x10^3) \tag{17}$$

and for radius $r$ in the magnetosphere [44], i.e. $2.873 \times 10^5$ m $> r - r_e > 10$ $r_e$:

$$\rho_{atm} = 1.6x10^{-17}10^{-0.285714\frac{r}{r_e}} \tag{18}$$

Mass density in Earth's magnetosphere depends strongly on solar activity and varies greatly. The data in Ref. 44 were averaged to produce equation (18), which should be adequate to estimate the annual event rate.

Our simulations show MQNs radiating between $10^{-28}$ W and $10^{+13}$ W and at frequencies between 0.03 MHz and ~2 GHz. Very high frequencies are associated with negligible RF power, and very high powers are associated with frequencies that are shielded by the magnetopause plasma. Results for RF power as a function of frequency are shown in Fig. 6 for events radiating at more than 1 µW and at frequencies more than 30 kHz. Results are shown for four representative and non-excluded values of $B_o$.

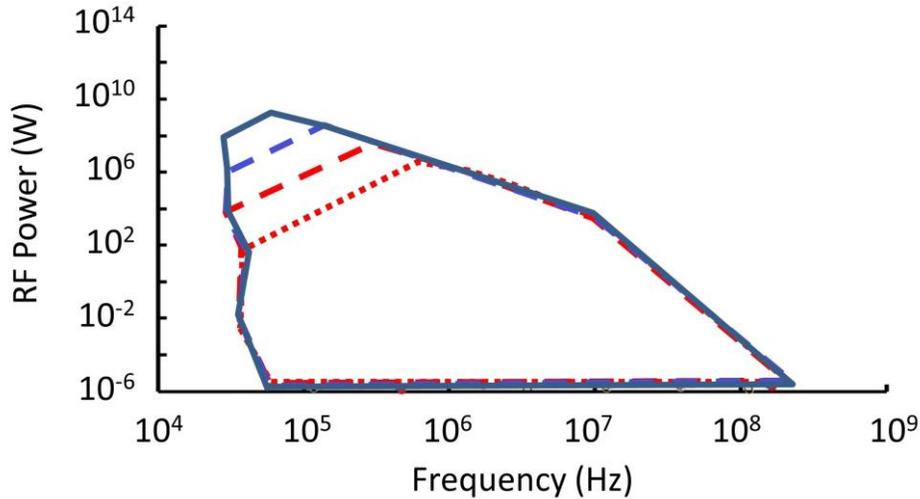

**Figure 6. Data points for RF power as a function of maximum equilibrium frequency for MQNs transiting through Earth's atmosphere for four representative values of $B_o$ are enclosed within the four perimeters: solid blue for $B_o = 3.0 \times 10^{12}$ T, dashed blue for $B_o = 2.5 \times 10^{12}$ T, dashed red for $B_o = 2.0 \times 10^{12}$ T, and dotted red for $B_o = 1.5 \times 10^{12}$ T.**

As shown in Fig. 6, highest power emissions occur at the lowest frequencies and highest values of $B_o$. These originate from the most massive MQNs penetrating the troposphere, but there are very few of them. The map associated with $B_o = 1.5 \times 10^{12}$ T is common to the maps of all $B_o$ values. The differences represent aggregation run-away as discussed in Ref. 4.

Figure 6 represents events by frequency and RF power but does not indicate the expected number of events per year. For each test MQN, the effective target area, given by equation (16), was multiplied by the corresponding number flux from Ref. 4 and by the 5.56 factor from equation (15), and summed over all simulated events with RF power greater than a sensor's detection threshold to estimate the number of events that might be observable per year as a function of detection threshold. The results are shown in Fig. 7.

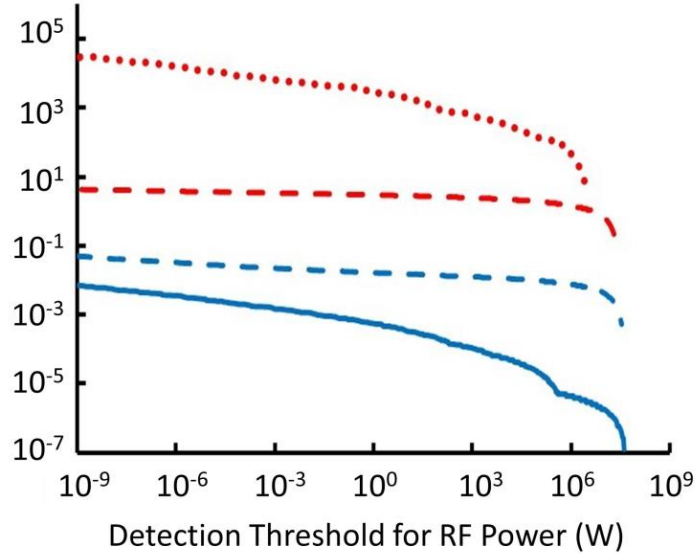

**Figure 7. Number of events per year expected, from all directions, above the indicated detection threshold of RF power for MQNs transiting through Earth's atmosphere for four representative values of $B_o$: solid blue for $B_o = 3.0 \times 10^{12}$ T, dashed blue for $B_o = 2.5 \times 10^{12}$ T, dashed red for $B_o = 2.0 \times 10^{12}$ T, and dotted red for $B_o = 1.5 \times 10^{12}$ T.**

At 1 nW threshold, the number of events per year that might be detectable in space out to 10 $r_e$ is ~30,000, ~8, ~0.8, and ~0.01 for $B_o = 1.5 \times 10^{12}$ T, $2.0 \times 10^{12}$ T, $2.5 \times 10^{12}$ T, and $3.0 \times 10^{12}$ T, respectively. For 1 µW detection threshold, the number events per year drops to ~10,000, ~7, ~0.07, and ~0.003 for $B_o = 1.5 \times 10^{12}$ T, $2.0 \times 10^{12}$ T, $2.5 \times 10^{12}$ T, and $3.0 \times 10^{12}$ T, respectively.

The number of events per year decreases so strongly with increasing $B_o$ because larger values of $B_o$ cause faster aggregation of MQNs in the early universe and, consequently, larger mass MQNs [4]. Since the mass per unit volume of dark matter is constrained by observations to be ~ $7 \times 10^{-22}$ kg/m³, the number density of MQNs decreases for increasing mass and increasing $B_o$. So the flux of MQNs and detection rate decrease for increasing $B_o$.

Detailed results for MQNs with RF power greater than 1 nW and with sufficient flux to be in the most probable 80% of events are provided in Supplementary Results: Representative Data Tables for Sensor Design. The information should be useful for designing sensors for detecting MQNs.

**Baseline sensor system**

A realistic sensor system is, of course, essential for testing the MQN dark-matter hypothesis. A convenient coincidence of frequency range and emerging technology enable a practical sensor. We outline one such baseline system and compute the corresponding event rate in this section.

As shown in Fig. 3, the system has three satellites equally spaced in a circular orbit with inclination 38.783° (to match MQN flux) at 51,000 km altitude where 1) the background plasma density is sufficiently low to permit good RF propagation in the intended detection band of $10^5$ to $10^6$ Hz, 2) radiation damage from electrons in the outer Van Allen belt is minimized, and 3) coverage by three satellites is acceptable. The Interplanetary Monitoring Platform IMP-6 (Explorer 43) spacecraft [45,46] is the reference architecture for the sensor system's spacecraft. IMP-6 was a 16-sided drum, 1.8-m long by 1.35-m diameter, having four 46-m long monopole antennas operating in pairs as 91 m long dipoles, and spinning at 5.4 revolutions per minute to scan space. Our baseline design is the same architecture but with 350 m long dipoles.

The sensor is based on the standard radar equation, formulated as transmit-receive equation:

$$P_r = G_r \frac{c^2}{4\pi f^2} G_s \frac{P_s}{4\pi R^2} \qquad (19)$$

where $P_r$ is the received power, $f$ is the frequency in Hz, $c$ is the speed of light, $G_r$ is the antenna gain, $P_s$ is the source power, $R$ is the distance, and $G_s$ is the source gain, which will be taken to be unity to represent the time averaged value. Lossless dipoles up to a half-wavelength long have gains $G_r < 1.6$ and directivities $D_r$ from 1.5 to 1.6. The weak signals require amplification in the receiver. The noise power $P_N$ and the signal to noise ratio $S/N$ for an amplified receiver are

$$P_N = \frac{G_r}{D_r} k_B T_{noise} \Delta f + k_B T_a \Delta f \qquad (20)$$

$$\frac{S}{N} = \left(\frac{c}{4\pi R f}\right)^2 \left(\frac{P_s}{k_B T_{noise} \Delta f / D_r + k_B T_a \Delta f / G_r}\right) \qquad (21)$$

where Boltzmann's constant $k_B = 1.38 \times 10^{-23}$ J/K, $T_{noise}$ is the absolute radiation noise temperature, $T_a$ is the preamplifier noise temperature, and $\Delta f$ is the receiver resolution bandwidth.

Solving equation (21) for the range $R$ gives

$$R = \frac{c}{4\pi f} \sqrt{\frac{G_r G_s P_s}{(S/N)}} = \frac{c}{4\pi f} \sqrt{\frac{P_s}{k_B T_{noise} \Delta f / D_r + k_B T_a \Delta f / G_r} \left(\frac{S}{N}\right)^{-1}}. \qquad (22)$$

Since many measurements will be made on each MQN as it transits through detection range, we can operate at signal/noise ratio $S/N = 1$ and use signal averaging and pattern recognition to reliably detect the signal.

The background noise temperature $T_{noise}$ is a major factor in determining sensor performance. IMP-6 measured the galactic background noise in the magnetosphere between 354 km and 206,000 km altitude and between 130 and 2,600 kHz. To determine noise temperatures from the noise power measurements of Brown [45], we equate his values for spectral brightness $B$ to the brightness in the Rayleigh-Jeans law for blackbody radiation and solve for the noise temperature $T_{noise}$.

$$B = \frac{2k_B f^2 T_{noise}}{c^2}, \text{ so } T_{noise} = \frac{c^2}{2k_B f^2} B \text{ or } T_{noise}(M^\circ K) = \frac{3.26 \times 10^{21}}{f^2 (MHz)^2} B \quad (23)$$

The results are shown in Table 1. The frequency range of interest encompasses a temperature maximum of 24 million degrees (MK) near 0.8 MHz. Ground based measurements at the poles during solar minima by Cane [47] are also shown in Table 1 for comparison and are consistent with our analysis of Brown's [45] data.

| F(MHz) | 0.2 | 0.3 | 0.5 | 0.8 | 1.0 | 2.0 | 3.0 |
|---|---|---|---|---|---|---|---|
| $T_{noise}$ (MK) | 3.1 | 6.2 | 21 | 24. | 19.5 | 8.2 | 4.0 |
| Cane [47] $T_{noise}$ (MK) | <7.3 | 11 | 15-38 | 18-27 | 16-23 | 6-11 | 2.6-5.0 |

**Table 1. Radiation temperature in millions of Kelvin degrees for various frequencies.**

Researchers accustomed to working at much higher frequencies, e.g. in satellite communications or radio astronomy, may find these values of $T_{noise}$ to be unreasonably high. However, $T_{noise}$ is decreasing from 0.8 MHz to 3.0 MHz in Table 1; it continues to decrease with increasing frequency to agree with $T_{noise}$ observed in those disciplines.

Fitting the data with the minimum of two polynomials reproduces the table for 0.2 MHz ≤ f ≤ 3.0 MHz within +/-1% and provides a useful function for estimating sensor performance:
$$T_{noise} = \min(659.03 f^4 - 1627.9 f^3 + 1313.5 f^2 - 359.49 f + 34.4, 40.0 f^{-2}) \text{ MK} \quad (24)$$

For these low frequencies $f$ (MHz), the noise temperature is millions of degrees Kelvin. From the denominator under the square-root term in equation (22), we see that variation of range with antenna gain is small if

$$G_r \geq D_r \frac{T_a}{T_{noise}}. \quad (25)$$

Since noise temperature $T_a$ can be less than 100 K (which does not require a cryogenic amplifier) for the frequencies of interest, 25 MK > $T_{noise}$ >3 MK in Table 1, and $D_r \sim 1.5$, equation (23) gives $6 \times 10^{-6} < G_r < 5 \times 10^{-5}$ before the range becomes very sensitive to antenna gain. For source powers of 1 μw, antenna gains of 0.001 to 1, radiation noise temperatures of 10 million K for the present case, amplifier noise temperatures of 100 K, receiver bandwidths of 1 Hz, and signal to noise ratios of unity, one gets ranges of 14,954 km and 15,740 km at 0.5 MHz. Therefore, range is very insensitive to antenna gain, and antenna gains that are well below unity are acceptable when the radiation temperature is very high.

The antenna gain $G_r$ depends on the dimensions and the termination impedance of the dipole antenna. Tang, Tieng, and Guna's theory [48] was used to design the antenna subsystem. The antenna gain $G_r$ as a function of frequency is shown in Fig. 8 for four terminating impedances.

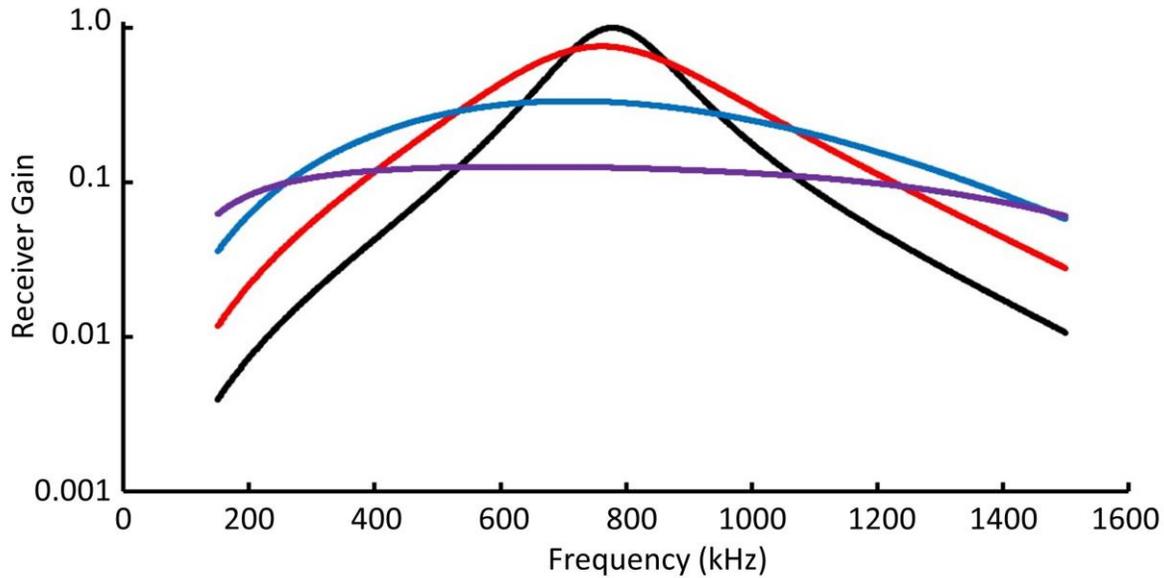

**Figure 8. Antenna gain is shown as a function of frequency for a load impedance of 65 Ω (—), 195 Ω (—), 650 Ω (—), and 1950 Ω (—). The matched impedance is 65 ohms.**

For a wide range of terminating impedances, $G_r$ easily satisfies equation (25) and the simple dipole is suitable for the baseline system. The optimal dipole design will likely incorporate resistive and/or possibly reactive loading at discrete positions along its length to achieve the best bandwidth and gain response if the additional complexity does not preclude deployment in space.

The other variables in equation (22) are source gain $G_s$ and signal-to-noise ratio (*S/N*). Both are set to 1.0. For multiple measurements like the satellites will be recording, detections with *S/N* = 1 are routine, and detections have been demonstrated for *S/N* as low as 0.1.

We found two options that apparently meet key requirements for the receiver: *Δf* = 1 Hz and a million simultaneously recorded frequencies. Assuming that such compact and low-power, ground-based electronic hardware can be made into space-qualified hardware, they provide proof-of-principle of receiver feasibility. The Tektronix RSA 306B Real Time Spectrum Analyzer system with Tektronix proprietary software on a separate computer monitors $10^6$ frequencies every 2 seconds with the required bandwidth Δf = 1 Hz and outputs the results to a laptop computer. In addition, the N210 USRP unit from National Instruments, which we have been using for autonomous data collection with our custom software for 254 frequencies every 25 μs, can meet the requirements by replacing the Field Programmable Gate Arrays (FPGAs) currently used for processing fast Fourier transforms with General Purpose Computing on Graphics Processing Units (GPGPUs).

In both cases, processing the data on $10^6$ simultaneous frequencies to find the unique signature of an MQN and extract the desired information will be a significant challenge. A combination of on-board processing with GPGPUs to select data for download and ground-based post-processing should be sufficient but has yet to be demonstrated.

**Estimated MQN detection rates**

Although the non-excluded range for $B_o$ at the time of this publication is $1 \times 10^{11}$ T $\leq B_o \leq 3 \times 10^{12}$ T, recent results being prepared for publication reduce the non-excluded range to $1.5 \times 10^{12}$ T $\leq B_o \leq 3 \times 10^{12}$ T. If $B_o < 1.5 \times 10^{12}$ T is included, the predicted MQN detection rate will be even larger and will shift to higher frequencies.

The number of events expected with the three-satellite sensor system described in this section was calculated for four values of $B_o$ in the more limited range. The calculation includes 1) computed number, frequency, and RF power of fly-through MQNs, 2) the measured RF background temperature as a function of frequency, 3) the sensor range as a function of MQN frequency and RF power from equation (22), and 4) the fraction of the area monitored by the three satellites at the 51,000-km altitude of their orbit as a function of MQN frequency and RF power. The number of MQNs that should be detected in five years of observations for with 0.1 MHz to 1.1 MHz RF is 1.2, 0.11, 0.005, or 0.0009 for $B_o$ = 1.5, 2.0, 2.5, or 3.0 Tera Tesla respectively. If the number of channels can be increased to 20 million so the band 0.1 MHz to 20.1 MHz can be monitored, then the expected number of events is increased by a factor of ~4.

Since quark nuggets are also baryons, their magnetic field may be similar to that of protons, which have the equivalent magnetic field $B_o$ between $0.9 \times 10^{12}$ T and $2.2 \times 10^{12}$ T. Calculated event rates assume the local mass density of dark matter equals the $7.0 \times 10^{-22}$ kg/m$^3$ of interstellar space. However, some MQNs should be aerocaptured by slowing down while passing through the Sun's corona, chromosphere, and photosphere and by subsequent gravitational interaction with a planet. These aerocaptured MQNs cannot return to the sun and should accumulate in the solar system to increase the event rate. The corresponding enhancement factor is currently unknown.

**Obtainable measurements and information**

An MQNs transiting through the magnetosphere, as shown in Fig. 3, provides a frequency source $f_0(t)$ moving at constant non-relativistic velocity $v_o$ in a straight line. A stationary observer at distance $r(t)$ records the observed frequency versus time. The source is a $x = v_o(t - t_o)$ for closest approach occurring at time $t_0$ and at distance $r_0$. For constant $df_o/dt$ rate of change of frequency, T\the observed frequency $f(t)$ at the receiver will be

$$f = \left( f_o + \frac{df_o}{dt}\left( t - \frac{\sqrt{x^2 + r_o^2}}{c} \right) \right)\left( 1 - \frac{v_o x}{c\sqrt{x^2 + r_o^2}} \right). \tag{26}$$

Total Doppler shifts are expected to be about that 1.5% of $f_0$. The source frequency is expected to decrease exponentially with time. During the brief periods of detectability, the source frequency should change linearly with time by amounts comparable to or less than the Doppler shifts.

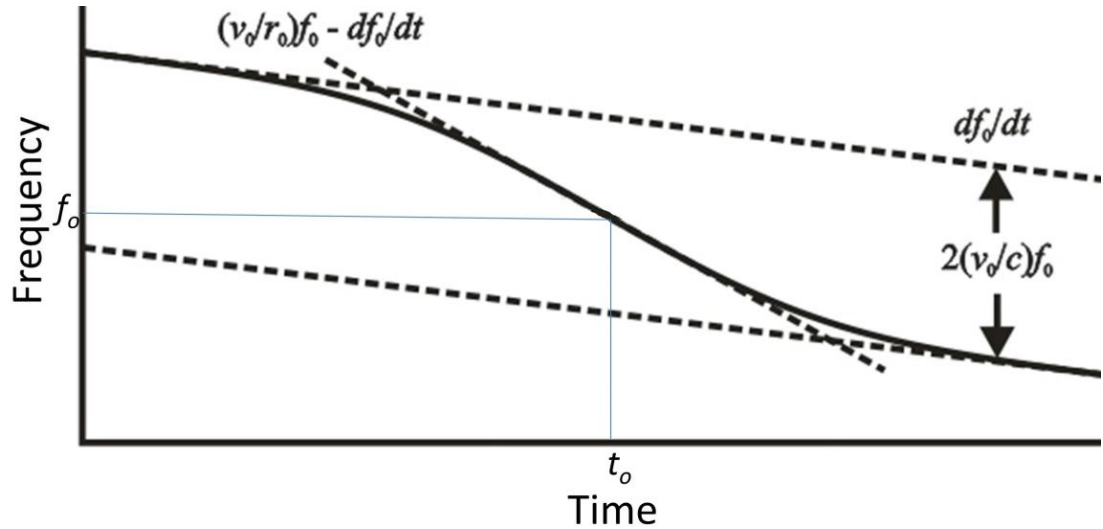

**Figure 9. Observed Doppler shifted frequency versus time for source radiating at $f_o$ = unshifted frequency, moving with $v_o$ = relative velocity, passing at $r_o$ = distance of closest approach at time $t_o$. Dotted lines show asymptotic rate of change of frequency $df_o/dt$ well before and after closest approach.**

Figure 9 shows what can be extracted from measurements of frequency versus time from the moving source. The velocity $v_0$ is determined from the asymptotic limit lines. The slope of these lines determines the rate of change of frequency of the source. The location of the inflection point determines $f_0$ and $t_0$, and the slope of the central asymptote determines $(v_0/r_0)f_0 - df_0/dt$ from which $r_0$ is determined with the other measured variables.

Because signal strengths are not likely to be measurable in the far asymptotic regions, it is important that good data be observed in the regions of maximum $d^2f/dt^2$, which occur at distances less than $2r_0$. Optimal fits of data to equation (26) will provide estimates of the variables of interest.

Thus, from the measurement it is possible to estimate $t_0$, $f_0$ $df_0/dt$, $r_0$, $v_0$, and $dv_0/dt$ and analyze the data to get two measurements of MQN mass, one from $f_0$ $df_0/dt$ and one from $v_0$, and $dv_0/dt$. With enough data the MQN mass distribution can be constructed and compared with the predictions on Ref. 4 and be used the refine Tatsumi's value of the surface magnetic field in magnetar cores.

**Discussion**

We have shown that MQNs transiting through ionized matter experience a torque on their magnetopause and spin up to high frequencies, ranging from kHz to GHz, depending on MQN mass and velocity, mass density of the surrounding matter, and the $B_o$ parameter. Rotating MQNs radiate. If the radiation is above the plasma frequency of the surrounding matter, the RF radiation will propagate and can, in principle, be used to detect MQN dark matter at a substantial distance.

We did examine the possibility of using ground-based sensors to detect MQNs transiting the magnetosphere and radiating at >40 MHz, the cut-off frequency of the ionosphere. The event rate is at most 0.3, 0.02, 0.0004, and 0.00007 per year even if $2\pi$ steradians solid angle can be observed. A space-based system is more promising.

Our results have identified requirements for MQN detection systems and a baseline system of three satellites that should test the MQN hypothesis for dark matter by detecting between 1,600 and 0.003 MQNs for $B_o$ between $1.5 \times 10^{12}$ T and $3.0 \times 10^{12}$ T, respectively. Recording their Doppler shifted frequencies during their passage by a sensor provides two different measurements of the MQN mass, so a mass distribution can be obtained in time.

The pattern of trajectories with respect to the direction of Vega, frequency, and rate of change of frequency would provide strong evidence of MQNs and eventually characterize the actual mass distribution of MQNs to compare with the predictions of Ref. 4.

Very low mass MQNs will have frequencies of up to 2 GHz, but their negligible RF power precludes their being detected. Very high mass MQNs will emit RF power of up to $10^{13}$ W, but their negligible flux makes their detection unlikely in a year of observation. As shown in Supplementary Results: Representative Data Tables for Sensor Design, frequencies between 100 kHz and 800 kHz are the most likely frequencies to be observed from MQNs.

Discerning quark-nugget signals from all human-caused and naturally occurring, non-MQN RF is facilitated by their characteristics: 1) narrow-band RF, unlike RF from lightning and other discharges, 2) continuous emissions, unlike pulsing radars, 3) relatively unmodulated in frequency and amplitude, unlike communication RF, 4) moving at ~200 km/s, unlike all human sources, and 5) initially increasing in frequency to a maximum and then slowly decreasing, unlike magnetosonic waves.

Finally, we note that solar and planetary atmospheres can be even larger-area, but less accessible, targets for RF-emitting MQNs.

**Data Availability**

All final analyzed data generated during this study are included in this published article with its Supplements.

**References**

1. Aghanim, N. *et al*. (Planck Collaboration), Planck 2018 results. VI. Cosmological parameters (2018). https://arxiv.org/pdf/1807.06209.pdf (2018). (Date of access: 10/05/2020).
2. Navarro, J. F.  Frenk, C. S. White, S. D. M. The structure of cold dark matter halos. *Astrophys. J*. **462,** 563-575 (1996).
3. Salucci, P. The distribution of dark matter in galaxies. *Astron Astrophys Rev* **27,** 2 (2019).
4. VanDevender, J. P., Shoemaker, I., Sloan T., VanDevender, A. P. & Ulmen, B.A. Mass distribution of magnetized quark-nugget dark matter and comparison with requirements and direct measurements. https://arxiv.org/abs/2004.12272 (2020). (Date of access: 10/05/2020).
5. Oerter, R. *The Theory of Almost Everything: The Standard Model, the Unsung Triumph of Modern Physics* (Penguin Group, 2006).
6. Witten, E. Cosmic separation of phases. *Phys. Rev. D* **30,** 272-285 (1984).
7. Farhi, E. & Jaffe, R. L. Strange matter. *Phys. Rev. D* **30,** 2379-2391 (1984).
8. De Rújula, A. & Glashow, S. L. Nuclearites—a novel form of cosmic radiation. *Nature* **312,** 734-737 (1984).
9. Zhitnitsky, A. "Nonbaryonic" dark matter as baryonic color superconductor. *JCAP* **0310,** 010 (2003).
10. Xia, C. J., Peng, G. X., Zhao, E. G. , and Zhou, S. G. From strangelets to strange stars: a unified description. *Sci. Bull.* **61,** 172 (2016).
11. Jacobs, D. M., Starkman, G. D., Lynn, B. W. Macro dark matter. *MNRAS* **450,** 3418–3430 (2015).
12. Wandelt, B. D. *et al.* Self-interacting dark matter. Ch. 5 in *Sources and Detection of Dark Matter and Dark Energy in the Universe*, from 4[th] Int'l*.* Symp.*,* Marina del Rey, CA, USA, February 23–25, 2000, (ed. Cline, D. B.) 263-274. (Springer, 2001). https://arxiv.org/abs/astro-ph/0006344, (2000)*.* (Date of access: 10/05/2020).
13. McCammon, D. *et al*. A high spectral resolution observation of the soft x-ray diffuse background with thermal detectors. *Astrophys. J.* **576,** 188 (2002).
14. Tulin, S. Self-Interacting dark matter. *AIP Conf. Proc.* **1604,** 121 (2014).
15. Tatsumi, T. Ferromagnetism of quark liquid. *Phys. Lett. B* **489,** 280-286 (2000).
16. VanDevender, J. P., *et al.* Detection of magnetized quark nuggets, a candidate for dark matter. *Sci. Rep*. **7,** 8758 (2017).
17. Lugones, G. & Horvath, J. E. Primordial nuggets survival and QCD pairing. *Phys. Rev. D* **69,** 063509 (2004).
18. Steiner, W. S., Reddy, S., & Prakash, M. Color-neutral superconducting dark matter. *Phys. Rev. D* **66,** 094007 (2002).
19. Bhattacharyya, A., *et al.* Relics of the cosmological QCD phase transition. *Phys. Rev. D* **61,** 083509 (2000).
20.  A. Chodos, *et al.* New extended model of hadrons. *Phys.Rev***. D9,** 12 3471-3495 (1974).
21. Aoki, Y., Endr, G., Fodor, Z., Katz, S. D. & Szabó, K. K. The order of the quantum chromodynamics transition predicted by the standard model of particle physics. *Nature* **443,** 675-678 (2006).
22. Bhattacharya, T. *et al*. QCD phase transition with chiral quarks and physical quark masses. *Phys. Rev. Lett.* **113,** 082001 (2014).
23. Gorham, P. W. and Rotter, B. J. Stringent neutrino flux constraints on anti-quark nugget dark matter. *Phys. Rev. D* **95,** 103002 (2017).
24. Ge, S., Lawson, K., & Zhitnitsky, A. The axion quark nugget dark matter model: size distribution and survival pattern. *Phys. Rev. D* **99,** 116017 (2019).
25. Atreya, A., Sarkar, A., & Srivastava, A. M. Reviving quark nuggets as a candidate for dark matter. *Phys. Rev. D* **90,** 045010 (2014).
26. Bazavov, A. *et al*. Additional strange hadrons from QCD thermodynamics and strangeness freeze out in heavy ion collisions. *Phys. Rev. Lett.* **113,** 072001 (2014).
26. Burdin, S., *et al*. Non-collider searches for stable massive particles. *Phys. Rep.* **582,** 1-52 (2015).
28. Chakrabarty, S. Quark matter in strong magnetic field. *Phys. Rev. D* **54,** 1306–1316 (1996).
29. Peng, G. X., Xu, J., & Xia, C-J. Magnetized strange quark matter in the equivparticle model with both confinement and perturbative interactions. *Nuc. Sci. Tech.* **27,** 98 (2016).
30. Patrignani, C., *et al.* Review of particle properties (Particle Data Group) *Chin. Phys.* **C40,** 100001 (2016).
31. Price, P. B. and Salamon, M. H. Search for supermassive magnetic monopoles using mica crystals. *Phys. Rev. Lett*. **56,** 12 1226-1229 (1986).
32. Porter. N. A., Fegan, D. J., MacNeill, G. C., Weekes, T. C. A search for evidence for nuclearites in astrophysical pulse experiments. *Nature* **316,** 49 (1985).
33. Porter. N. A., Cawley, M. F., Fegan, D. J., MacNeill, G. C., Weekes, T. C. A search for evidence for nuclearites in astrophysical pulse experiments. *Irish Astron. J*. **18,** 193-196 (1988).

34. Bassan, M. *et al.* Dark matter searches using gravitational wave bar detectors: quark nuggets and nuclearites. *Astropart. Phys.* **78,** 52-64 (2016).
35. Scherrer, R. J. & Turner, M. S. On the relic, cosmic abundance of stable, weakly interacting massive particles. *Phys. Rev. D* **33,** 1585-1589 (1986).
36. Spitzer, L. *Physics of fully ionized gases*, 2nd edition, 4-22 (Wiley Interscience, 1962).
37. Papagiannis, M. D. The torque applied by the solar wind on the tilted magnetosphere. *J. Geophys. Res.* **78,** 34, 7968–7977 (1973).
38. Jackson, J. D. *Classical Electrodynamics*, 3rd edition, 413-414 (Wiley and Sons (Asia), 1999).
39. Harrison, E. R. Olbers' paradox and the background radiation density in an isotropic homogeneous universe. *Mon. Not. R. astr. Soc.* **131,** 1-12 (1965).
40. Ferriere, K. The interstellar environment of our galaxy. *Rev. Mod. Phys.* **73,** 1031-1066 (2001).
41. Vogelsberger, M. & Zavala, J. Direct detection of self-interacting dark matter. *MNRAS* **430,** 1722–1735 (2013).
42. Cai, C., Khasawneh, K., Liu, H., Wei, M. Collisionless Gas Flows over a Cylindrical or Spherical Object. *J. Spacecraft Rockets* **46,** 1124–1131 (2009).
43. Havens, R. J., Koll, R. T., and LaGow, H. E. The pressure, density, and temperature of the Earth's atmosphere to 160 kilometers. *J. Geophys. Res.* **57,** 1 59-72 (1952).
44. Denton, R. E., Menietti, J. D., Goldstein, J., Young, S. L. and Anderson R. R. Electron density in the magnetosphere. *J. Geophys. Res.* **109,** A09215 (2004).
45. Brown, L. W. The galactic radio spectrum between 130 and 2600 kHz. *Astrophys. J.* **180,** 359-370 (1973).
46. Frankel, M. S. LF radio noise from Earth's magnetosphere. *Radio Sci.* **8,** 11, 991-1005 (1973).
47. Cane, H. V. Spectra of the non-thermal radio radiation from the galactic polar regions. *Mon. Not. R. astr. Soc.* **189,** 465-478 (1979).
48. Tang, T. G., Tieng, Q. M., Gunn, M. W. Equivalent circuit of a dipole antenna using frequency-independent lumped elements. *IEEE T. Antenn. Propag.* **41**, 1, 100-103 (1993).

**Figure Captions**

**Figure 1. Generalization of the tan*(χ)* factor in equation (5), in which *χ* is the angle between the magnetic axis and the normal to both the magnetic axis and the quark-nugget's direction of travel in the rest frame of the quark nugget.** The solid red lines indicate the angles computed by Papagiannis; the solid blue lines indicate extensions by symmetry. Dotted blue lines indicate functional extrapolation of Papagiannis and symmetry-extension values.

**Figure 2. Calculated frequency versus time for 0.1 kg MQN with $B_o = 2.25 \times 10^{12}$ T, initial velocity of 250 km/s, and passing through air at density 1.0 kg/m$^3$.** Note the initial oscillation is about 0 until angular momentum becomes sufficient to complete a full rotation.

**Figure 3. Near Earth environment with MQN trajectories (red lines).** Three concentric circles represent MQN interaction volume: highly ionized and low density magnetosphere and ionosphere (purple); weakly ionized or neutral, low-density troposphere (blue), and neutral, high-density planet (gray). Satellites S1, S2 and S3 are shown in orbits that let them monitor narrow-band RF emissions above background as described below. The vertical ring (light blue) is a detection-area element for simulations of MQNs interacting with magnetosphere, ionosphere and troposphere, as discussed below.

**Figure 4. For the first day of each month, Earth's position and velocity about the Sun are shown.** The solar system's velocity towards Vega is shown by the black vector from the Sun. The net velocity vector of dark matter into Earth is shown in blue for each month. The effects of the 23.5° angle between Earth's equatorial plane and the ecliptic and the 38.8° angle between Earth's equatorial plane and Vega's position are not shown.

**Figure 5.** Calculated and normalized variation of dark matter flux streaming from the direction of the star Vega as a function of polar angle $\theta$ from Vega's zenith.

**Figure 6.** Data points for RF power as a function of maximum equilibrium frequency for MQNs transiting through Earth's atmosphere for four representative values of $B_o$ are enclosed within the four perimeters: solid blue for $B_o = 3.0 \times 10^{12}$ T, dashed blue for $B_o = 2.5 \times 10^{12}$ T, dashed red for $B_o = 2.0 \times 10^{12}$ T, and dotted red for $B_o = 1.5 \times 10^{12}$ T.

**Figure 7.** Number of events per year expected, from all directions, above the indicated detection threshold of RF power for MQNs transiting through Earth's atmosphere for four representative values of $B_o$: solid blue for $B_o = 3.0 \times 10^{12}$ T, dashed blue for $B_o = 2.5 \times 10^{12}$ T, dashed red for $B_o = 2.0 \times 10^{12}$ T, and dotted red for $B_o = 1.5 \times 10^{12}$ T.

**Figure 8.** Antenna gain is shown as a function of frequency for a load impedance of 65 Ω (—), 195 Ω (—), 650 Ω (—), and 1950 Ω (—). The matched impedance is 65 ohms.

**Figure 9.** Observed Doppler shifted frequency versus time for source radiating at $f_o$ = unshifted frequency, moving with $v_o$ = relative velocity, passing at $r_o$ = distance of closest approach at time $t_o$. Dotted lines show asymptotic rate of change of frequency $df_o/dt$ well before and after closest approach.

### Acknowledgements

We gratefully acknowledge S. V. Greene for first suggesting that quark nuggets might explain the geophysical evidence that initiated this research (she generously declined to be a coauthor), Albuquerque Academy for hosting preliminary experiments, Robert Nellums for reviewing portions of the work, and Jesse A. Rosen for editing and improving this paper.

This work was supported by VanDevender Enterprises, LLC.

### Author Contributions

J.P.V. was lead physicist and principal investigator. He developed computer programs to calculate the quark-nugget mass distribution, analyzed the results, wrote the paper, prepared the figures, and revised the paper to incorporate the improvements from the other author and reviewers.

C. J. B. analyzed the detectability of the RF emissions from MQNs and provided valuable comments on the paper.

C. C. analyzed the expected distribution of signals as a function of time of year, time of day, and sensor position with a realistic MQN velocity distribution, including the motion of Earth through the dark-matter halo towards the star Vega.

A. P. V. red-teamed (provided critical, independent, review of) the analysis that avoided unjustified or erroneous conclusions.

B. A. U. contributed important suggestions on planets to include and discussion of the relevant plasma physics of the magnetopause interaction.

**Additional Information**

Competing Financial Interests

The authors declare that there are no competing financial interests.

Supplementary Information for

**Radio frequency emissions from dark-matter-candidate magnetized quark nuggets interacting with matter**

J. Pace VanDevender*, C. Jerald Buchenauer, Chunpei Cai, Aaron P. VanDevender, and Benjamin A. Ulmen
*Corresponding author at pace@vandevender.com

**Supplementary Note: Quark-nugget research summary**

Witten [6] showed quark-nuggets are in the theoretically predicted, ultra-dense, color-flavor-locked (CFL) phase [17] of quark matter. Steiner, *et al.* [18] showed that the ground state of the CFL phase is color neutral and that color neutrality forces electric charge neutrality, which minimizes electromagnetic emissions. However, Xia, *et al.* [10] found that quark depletion causes the ratio $Q/A$ of electric charge $Q$ to baryon number $A$ to be non-zero and varying at $Q/A \sim 0.32\, A^{-1/3}$ for $3 < A < 10^5$. In addition to this core charge, they find that there is a large surface charge and a neutralizing cloud of charge to give a net zero electric charge for sufficiently large $A$. So quark nuggets with $A \gg 1$ are both dark and very difficult to detect with astrophysical observations.

Witten and Xia, *et al.* also showed their density should be somewhat larger than the density of nuclei, and their mass very large, even the mass of a star. Large quark nuggets are predicted to be stable [6, 7, 17-19] with mass between $10^{-8}$ kg and $10^{20}$ kg within a plausible but uncertain range of assumed parameters of quantum chromodynamics (QCD) and the MIT bag model with its inherent limitations [20].

Although Witten assumed a first-order phase transition formed quark nuggets, Aoki, *et al.*[21] showed that the finite-temperature QCD transition that formed quark nuggets in the hot early universe was very likely an analytic crossover, involving a rapid change as the temperature varied, but not a real phase transition. Recent simulations by T. Bhattacharya, *et al*. [22] support the crossover process.

A combination of quark nuggets and anti-quark nuggets have also been proposed within constraints imposed by observations of neutrino flux [23]. Zhitnitsky [9] proposed that Axion Quark Nuggets (AQN) that forms quark and anti-quark nuggets were generated by the collapse of the axion domain wall network. Although the model relies on the hypothetical particle that is a proposed extension of the Standard Model to explain CP violation, it appears to explain a wide variety of longstanding problems and leads to quark and anti-quark nuggets with a narrow mass distribution at ~10 kg [24]. Atreya, *et al*. [25] also found that CP-violating quark and anti-quark scatterings from moving Z(3) domain walls should form quark and anti-quark nuggets, regardless of the order of the quark-hadron phase transition.

Experiments by A. Bazavov, *et al*. [26] at the Relativistic Heavy Ion Collider (RHIC) have provided the first indirect evidence of strange baryonic matter. Additional experiments at RHIC may determine whether the process is a first order phase transition or the crossover process. In either case, quark nuggets could have theoretically formed in the early universe.

In 2001, Wandelt, *et al*. [12] showed that quark nuggets meet all the theoretical requirements for dark matter and are not excluded by observations when the stopping power for quark nuggets in the materials covering a detector is properly considered and when the average mass is >$10^5$ GeV (~$2 \times 10^{-22}$ kg). In 2014, Tulin [14] surveyed additional simulations of increasing sophistication and updated the results of Wandelt, *et al*. The combined results help establish the allowed range and velocity dependence of the strength parameter and strengthen the case for quark nuggets. In 2015, Burdin, *et al.* [27] examined all non-accelerator candidates for stable dark matter and also concluded that quark nuggets meet the requirements for dark matter and have not been excluded experimentally. Jacobs, Starkman, and Lynn [11] found that combined Earth-based, astrophysical, and cosmological observations still allow quark nuggets of mass 0.055 to $10^{14}$ kg and $2 \times 10^{17}$ to $4 \times 10^{21}$ kg to contribute substantially to dark matter. The large mass means the number per unit volume of space is small, so detecting them requires a very large-area detector.

These studies did not consider an intrinsic magnetic field within quark nuggets. However, Tatsumi [15] has shown that the lowest-energy configuration of a quark nugget depends on the QCD coupling constant and can be a ferromagnetic liquid that can account for magnetars. He calculates the value of the magnetic field at the surface of a quark-nugget core inside a magnetar to be $10^{12\pm1}$ T, which is large compared to expected values for the magnetic field at the surface of a magnetar star with a quark-nugget core. For a quark nugget of radius $r_{QN}$ and a magnetar of radius $r_s$, the magnetic field scales as $(r_{QN}/r_s)^3$. Therefore, the surface magnetic field of a magnetar is smaller than $10^{12}$ T because $r_s > r_{QN}$. Since quark-nugget dark matter is bare, the surface magnetic field of what we wish to detect is $10^{12\pm1}$ T.

Although the cross section for interacting with dense matter is greatly enhanced [16] by the magnetic field which falls off as radius $r_{QN}^{-3}$, the collision cross section is still many orders of magnitude too small to violate the collision requirements [11, 12, 14, 27] for dark matter and will be discussed below.

Chakrabarty [28] showed that the stability of quark nuggets increases with increasing external magnetic field $\leq 10^{16}$ T, so the large self-field described by Tatsumi should enhance their stability. Ping, *et al.* [29] showed that magnetized quark nuggets should be absolutely stable with the newly-developed equivparticle model, so the large self-field described by Tatsumi should ensure that quark nuggets with sufficiently large baryon number will not decay by the weak interaction.

The large magnetic field also alters MQN interaction with ordinary matter through the greatly-enhanced stopping power of the magnetopause around high-velocity MQNs moving through a plasma [16]. Searches [30] for quark nuggets with underground detectors would not be sensitive to highly magnetized quark nuggets, which cannot penetrate the material above the detector. For example, the paper by Gorham and Rotter [23] about constraints on anti-quark nugget dark matter (which do not constrain quark-nuggets unless the ratio of anti-quark nuggets to quark nuggets is shown to be large) assumes that limits on the flux of magnetic monopoles from analysis by Price, *et al.* [31] of geologic mica buried under 3 km of rock are also applicable to quark nuggets. Gorham and Rotter also cite work by Porter, *et al.* [32-33] as constraining quark-nugget (nuclearite) contributions to dark matter by the absence of meteor-like objects in the lower atmosphere that are fast enough to be quark nuggets. Bassan, *et al*. [34] looked for quark nuggets (nuclearites) with gravitational wave detectors and found signals much less than expected for the flux of dark matter. However, all of these analyses assumed quark nuggets can reach the detector volume because the cross section for momentum transfer is the geometric

-28-

cross section. In contrast, the MQN magnetopause cross section [16] is many orders of magnitude larger and prevents all but the most massive MQNs from being detected.

**Supplementary Results: Representative Data Tables for Sensor Design**

Table S1 through Table S4 respectively provide representative events for $B_o$ equals $1.5 \times 10^{12}$ T, $2.0 \times 10^{12}$ T, $2.5 \times 10^{12}$ T, and $3.0 \times 10^{12}$ T. The tables show the computed parameters for MQN events with RF power greater than 1 nW and with sufficient flux to place them in the most probable 80% of the total number of events. The information should be useful for designing sensors for detecting MQNs.

The computed mass distribution [4] of MQNs extend over 30 orders of magnitude, from ~ $10^{-24}$ kg to > $10^6$ kg. To cover such a large range, we approximated the mass distributions by the distribution of decadal masses. Calculated trajectories for the masses in the logarithmic center of each decade of mass approximate the behaviors of all the MQNs in that decade. Decade mass is listed in the left most column of the tables. The total flux for all masses in a decade of mass have been calculated in aggregation simulations [4] and that decadal flux is given in the 7th column of the Tables. Multiplying that flux by the cross section corresponding to the closest-approach altitude $h$ and extent $\Delta h$ for the entry and the 5.56 factor to generalize the unidirectional result to omni-directional results gives the event rate in the 8th column. The spin-down time is shown in the 5th column. Each of the four tables is for the indicated value of the surface-magnetic-field parameter $B_o$.

**Table S1: Representative results for $B_o = 1.5 \times 10^{12}$ T.**

| MQN Mass (kg) | Altitude h (m) | Delta_h (m) | Frequency (Hz) | $\tau\_down$ (s) | RF power (W) | Flux (n/m^2/y/sr) | Number/y for all directions |
|---|---|---|---|---|---|---|---|
| $3.00\times10^{-8}$ | $1.03\times10^{5}$ | $1.03\times10^{4}$ | $2.61\times10^{8}$ | $5.97\times10^{1}$ | $1.00\times10^{-9}$ | $2.12\times10^{-13}$ | $3.16\times10^{3}$ |
| $3.00\times10^{-8}$ | $9.28\times10^{4}$ | $9.28\times10^{3}$ | $3.77\times10^{8}$ | $2.86\times10^{1}$ | $4.38\times10^{-9}$ | $2.12\times10^{-13}$ | $2.58\times10^{3}$ |
| $3.00\times10^{-8}$ | $8.36\times10^{4}$ | $8.36\times10^{3}$ | $4.89\times10^{8}$ | $1.70\times10^{1}$ | $1.23\times10^{-8}$ | $2.12\times10^{-13}$ | $2.12\times10^{3}$ |
| $3.00\times10^{-5}$ | $9.28\times10^{4}$ | $9.28\times10^{3}$ | $4.06\times10^{7}$ | $2.47\times10^{2}$ | $5.88\times10^{-7}$ | $1.20\times10^{-13}$ | $1.79\times10^{3}$ |
| $3.00\times10^{-6}$ | $1.03\times10^{5}$ | $1.03\times10^{4}$ | $5.81\times10^{7}$ | $2.60\times10^{2}$ | $2.47\times10^{-8}$ | $1.16\times10^{-13}$ | $1.73\times10^{3}$ |
| $3.00\times10^{-5}$ | $8.36\times10^{4}$ | $8.36\times10^{3}$ | $5.87\times10^{7}$ | $1.18\times10^{2}$ | $2.57\times10^{-6}$ | $1.20\times10^{-13}$ | $1.47\times10^{3}$ |
| $3.00\times10^{-6}$ | $9.28\times10^{4}$ | $9.28\times10^{3}$ | $8.65\times10^{7}$ | $1.17\times10^{2}$ | $1.21\times10^{-7}$ | $1.16\times10^{-13}$ | $1.41\times10^{3}$ |
| $3.00\times10^{-5}$ | $7.52\times10^{4}$ | $7.52\times10^{3}$ | $7.99\times10^{7}$ | $6.38\times10^{1}$ | $8.81\times10^{-6}$ | $1.20\times10^{-13}$ | $1.20\times10^{3}$ |
| $3.00\times10^{-6}$ | $8.36\times10^{4}$ | $8.36\times10^{3}$ | $1.23\times10^{8}$ | $5.84\times10^{1}$ | $4.88\times10^{-7}$ | $1.16\times10^{-13}$ | $1.16\times10^{3}$ |
| $3.00\times10^{-7}$ | $1.03\times10^{5}$ | $1.03\times10^{4}$ | $1.24\times10^{8}$ | $1.23\times10^{2}$ | $5.08\times10^{-9}$ | $7.40\times10^{-14}$ | $1.10\times10^{3}$ |
| $3.00\times10^{-5}$ | $6.77\times10^{4}$ | $6.77\times10^{3}$ | $1.03\times10^{8}$ | $3.80\times10^{1}$ | $2.48\times10^{-5}$ | $1.20\times10^{-13}$ | $9.85\times10^{2}$ |
| $3.00\times10^{-6}$ | $7.52\times10^{4}$ | $7.52\times10^{3}$ | $1.64\times10^{8}$ | $3.28\times10^{1}$ | $1.55\times10^{-6}$ | $1.16\times10^{-13}$ | $9.51\times10^{2}$ |
| $3.00\times10^{-7}$ | $9.28\times10^{4}$ | $9.28\times10^{3}$ | $1.82\times10^{8}$ | $5.67\times10^{1}$ | $2.40\times10^{-8}$ | $7.40\times10^{-14}$ | $9.04\times10^{2}$ |
| $3.00\times10^{-5}$ | $6.09\times10^{4}$ | $6.09\times10^{3}$ | $1.26\times10^{8}$ | $2.55\times10^{1}$ | $5.50\times10^{-5}$ | $1.20\times10^{-13}$ | $8.09\times10^{2}$ |
| $3.00\times10^{-6}$ | $6.77\times10^{4}$ | $6.77\times10^{3}$ | $2.02\times10^{8}$ | $2.16\times10^{1}$ | $3.57\times10^{-6}$ | $1.16\times10^{-13}$ | $7.81\times10^{2}$ |
| $3.00\times10^{-7}$ | $8.36\times10^{4}$ | $8.36\times10^{3}$ | $2.51\times10^{8}$ | $3.00\times10^{1}$ | $8.57\times10^{-8}$ | $7.40\times10^{-14}$ | $7.41\times10^{2}$ |
| $3.00\times10^{-3}$ | $7.52\times10^{4}$ | $7.52\times10^{3}$ | $1.78\times10^{7}$ | $2.77\times10^{2}$ | $2.16\times10^{-4}$ | $4.46\times10^{-14}$ | $6.65\times10^{2}$ |
| $3.00\times10^{-7}$ | $7.52\times10^{4}$ | $7.52\times10^{3}$ | $3.15\times10^{8}$ | $1.90\times10^{1}$ | $2.14\times10^{-7}$ | $7.40\times10^{-14}$ | $6.08\times10^{2}$ |
| $3.00\times10^{-3}$ | $6.77\times10^{4}$ | $6.77\times10^{3}$ | $2.40\times10^{7}$ | $1.53\times10^{2}$ | $7.14\times10^{-4}$ | $4.46\times10^{-14}$ | $5.44\times10^{2}$ |
| $3.00\times10^{-4}$ | $8.36\times10^{4}$ | $8.36\times10^{3}$ | $2.73\times10^{7}$ | $2.54\times10^{2}$ | $1.20\times10^{-5}$ | $3.51\times10^{-14}$ | $5.23\times10^{2}$ |
| $3.00\times10^{-3}$ | $6.09\times10^{4}$ | $6.09\times10^{3}$ | $3.07\times10^{7}$ | $9.28\times10^{1}$ | $1.93\times10^{-3}$ | $4.46\times10^{-14}$ | $4.46\times10^{2}$ |
| $3.00\times10^{-4}$ | $7.52\times10^{4}$ | $7.52\times10^{3}$ | $3.79\times10^{7}$ | $1.32\times10^{2}$ | $4.45\times10^{-5}$ | $3.51\times10^{-14}$ | $4.29\times10^{2}$ |
| $3.00\times10^{-1}$ | $4.93\times10^{4}$ | $4.93\times10^{3}$ | $1.07\times10^{7}$ | $1.65\times10^{2}$ | $2.84\times10^{-1}$ | $2.63\times10^{-14}$ | $3.93\times10^{2}$ |
| $3.00\times10^{-3}$ | $5.48\times10^{4}$ | $5.48\times10^{3}$ | $3.83\times10^{7}$ | $5.99\times10^{1}$ | $4.63\times10^{-3}$ | $4.46\times10^{-14}$ | $3.66\times10^{2}$ |
| $3.00\times10^{-4}$ | $6.77\times10^{4}$ | $6.77\times10^{3}$ | $5.06\times10^{7}$ | $7.39\times10^{1}$ | $1.41\times10^{-4}$ | $3.51\times10^{-14}$ | $3.51\times10^{2}$ |
| $3.00\times10^{-1}$ | $4.44\times10^{4}$ | $4.44\times10^{3}$ | $1.29\times10^{7}$ | $1.13\times10^{2}$ | $6.05\times10^{-1}$ | $2.63\times10^{-14}$ | $3.21\times10^{2}$ |
| $3.00\times10^{-3}$ | $4.93\times10^{4}$ | $4.93\times10^{3}$ | $4.58\times10^{7}$ | $4.19\times10^{1}$ | $9.49\times10^{-3}$ | $4.46\times10^{-14}$ | $3.00\times10^{2}$ |
| $3.00\times10^{-4}$ | $6.09\times10^{4}$ | $6.09\times10^{3}$ | $6.36\times10^{7}$ | $4.68\times10^{1}$ | $3.53\times10^{-4}$ | $3.51\times10^{-14}$ | $2.88\times10^{2}$ |
| $3.00\times10^{-2}$ | $6.09\times10^{4}$ | $6.09\times10^{3}$ | $1.46\times10^{7}$ | $1.92\times10^{2}$ | $9.75\times10^{-3}$ | $1.93\times10^{-14}$ | $2.88\times10^{2}$ |
| $3.00$ | $3.60\times10^{4}$ | $3.60\times10^{3}$ | $8.43\times10^{6}$ | $1.23\times10^{2}$ | $1.09\times10^{1}$ | $1.93\times10^{-14}$ | $2.88\times10^{2}$ |
| $3.00\times10^{-1}$ | $4.00\times10^{4}$ | $4.00\times10^{3}$ | $1.52\times10^{7}$ | $8.21\times10^{1}$ | $1.14$ | $2.63\times10^{-14}$ | $2.63\times10^{2}$ |
| $3.00\times10^{-3}$ | $4.44\times10^{4}$ | $4.44\times10^{3}$ | $5.31\times10^{7}$ | $3.11\times10^{1}$ | $1.72\times10^{-2}$ | $4.46\times10^{-14}$ | $2.47\times10^{2}$ |
| $3.00\times10^{-4}$ | $5.48\times10^{4}$ | $5.48\times10^{3}$ | $7.68\times10^{7}$ | $3.20\times10^{1}$ | $7.51\times10^{-4}$ | $3.51\times10^{-14}$ | $2.37\times10^{2}$ |
| $3.00\times10^{-2}$ | $5.48\times10^{4}$ | $5.48\times10^{3}$ | $1.85\times10^{7}$ | $1.19\times10^{2}$ | $2.54\times10^{-2}$ | $1.93\times10^{-14}$ | $2.36\times10^{2}$ |

**Table S1: Representative results for $B_o = 2.0 \times 10^{12}$ T.**

| MQN Mass (kg) | Altitude $h$ (m) | $\Delta h$ (m) | Frequency (Hz) | $\tau\_down$ (s) | RF power (W) | Flux ($m^{-2}y^{-1}sr^{-1}$) | Number/y for all directions |
|---|---|---|---|---|---|---|---|
| $3.00\times10^6$ | $6.14\times10^3$ | $6.14\times10^2$ | $2.81\times10^5$ | $6.27\times10^2$ | $2.38\times10^7$ | $1.46\times10^{-16}$ | $1.06\times10^{-1}$ |
| $3.00\times10^6$ | $7.14\times10^3$ | $7.14\times10^2$ | $2.70\times10^5$ | $6.79\times10^2$ | $2.03\times10^7$ | $1.46\times10^{-16}$ | $1.06\times10^{-1}$ |
| $3.00\times10^6$ | $8.14\times10^3$ | $8.14\times10^2$ | $2.59\times10^5$ | $7.35\times10^2$ | $1.73\times10^7$ | $1.46\times10^{-16}$ | $1.06\times10^{-1}$ |
| $3.00\times10^6$ | $9.14\times10^3$ | $9.14\times10^2$ | $2.49\times10^5$ | $7.96\times10^2$ | $1.48\times10^7$ | $1.46\times10^{-16}$ | $1.06\times10^{-1}$ |
| $3.00\times10^6$ | $1.02\times10^4$ | $1.02\times10^3$ | $2.39\times10^5$ | $8.62\times10^2$ | $1.26\times10^7$ | $1.46\times10^{-16}$ | $1.06\times10^{-1}$ |
| $3.00\times10^6$ | $1.13\times10^4$ | $1.13\times10^3$ | $2.28\times10^5$ | $9.52\times10^2$ | $1.03\times10^7$ | $1.46\times10^{-16}$ | $1.06\times10^{-1}$ |
| $3.00\times10^6$ | $1.25\times10^4$ | $1.25\times10^3$ | $2.17\times10^5$ | $1.05\times10^3$ | $8.46\times10^6$ | $1.46\times10^{-16}$ | $1.06\times10^{-1}$ |
| $3.00\times10^6$ | $1.39\times10^4$ | $1.39\times10^3$ | $2.06\times10^5$ | $1.16\times10^3$ | $6.93\times10^6$ | $1.46\times10^{-16}$ | $1.06\times10^{-1}$ |
| $3.00\times10^6$ | $1.55\times10^4$ | $1.55\times10^3$ | $1.94\times10^5$ | $1.31\times10^3$ | $5.46\times10^6$ | $1.46\times10^{-16}$ | $1.06\times10^{-1}$ |
| $3.00\times10^6$ | $1.72\times10^4$ | $1.72\times10^3$ | $1.81\times10^5$ | $1.50\times10^3$ | $4.13\times10^6$ | $1.46\times10^{-16}$ | $1.06\times10^{-1}$ |
| $3.00\times10^6$ | $1.91\times10^4$ | $1.91\times10^3$ | $1.67\times10^5$ | $1.76\times10^3$ | $3.01\times10^6$ | $1.46\times10^{-16}$ | $1.06\times10^{-1}$ |
| $3.00\times10^6$ | $2.12\times10^4$ | $2.12\times10^3$ | $1.53\times10^5$ | $2.11\times10^3$ | $2.10\times10^6$ | $1.46\times10^{-16}$ | $1.06\times10^{-1}$ |
| $3.00\times10^6$ | $2.36\times10^4$ | $2.36\times10^3$ | $1.40\times10^5$ | $2.52\times10^3$ | $1.47\times10^6$ | $1.46\times10^{-16}$ | $1.06\times10^{-1}$ |
| $3.00\times10^6$ | $2.62\times10^4$ | $2.62\times10^3$ | $1.25\times10^5$ | $3.14\times10^3$ | $9.48\times10^5$ | $1.46\times10^{-16}$ | $1.06\times10^{-1}$ |
| $3.00\times10^6$ | $2.91\times10^4$ | $2.91\times10^3$ | $1.12\times10^5$ | $3.91\times10^3$ | $6.12\times10^5$ | $1.46\times10^{-16}$ | $1.06\times10^{-1}$ |
| $3.00\times10^6$ | $3.24\times10^4$ | $3.24\times10^3$ | $9.77\times10^4$ | $5.17\times10^3$ | $3.50\times10^5$ | $1.46\times10^{-16}$ | $1.06\times10^{-1}$ |
| $3.00\times10^6$ | $3.60\times10^4$ | $3.60\times10^3$ | $8.50\times10^4$ | $6.83\times10^3$ | $2.01\times10^5$ | $1.46\times10^{-16}$ | $1.06\times10^{-1}$ |
| $3.00\times10^6$ | $4.00\times10^4$ | $4.00\times10^3$ | $7.25\times10^4$ | $9.39\times10^3$ | $1.06\times10^5$ | $1.46\times10^{-16}$ | $1.06\times10^{-1}$ |
| $3.00\times10^6$ | $4.44\times10^4$ | $4.44\times10^3$ | $6.06\times10^4$ | $1.34\times10^4$ | $5.19\times10^4$ | $1.46\times10^{-16}$ | $1.06\times10^{-1}$ |
| $3.00\times10^6$ | $4.93\times10^4$ | $4.93\times10^3$ | $4.97\times10^4$ | $2.00\times10^4$ | $2.34\times10^4$ | $1.46\times10^{-16}$ | $1.06\times10^{-1}$ |
| $3.00\times10^6$ | $5.48\times10^4$ | $5.48\times10^3$ | $3.99\times10^4$ | $3.10\times10^4$ | $9.75\times10^3$ | $1.46\times10^{-16}$ | $1.06\times10^{-1}$ |
| $3.00\times10^6$ | $6.09\times10^4$ | $6.09\times10^3$ | $3.11\times10^4$ | $5.09\times10^4$ | $3.60\times10^3$ | $1.46\times10^{-16}$ | $1.06\times10^{-1}$ |
| $3.00\times10^6$ | $6.77\times10^4$ | $6.77\times10^3$ | $2.38\times10^4$ | $8.72\times10^4$ | $1.23\times10^3$ | $1.46\times10^{-16}$ | $1.06\times10^{-1}$ |
| $3.00\times10^6$ | $7.52\times10^4$ | $7.52\times10^3$ | $1.75\times10^4$ | $1.62\times10^5$ | $3.58\times10^2$ | $1.46\times10^{-16}$ | $1.06\times10^{-1}$ |
| $3.00\times10^6$ | $8.36\times10^4$ | $8.36\times10^3$ | $1.26\times10^4$ | $3.12\times10^5$ | $9.63\times10^1$ | $1.46\times10^{-16}$ | $1.06\times10^{-1}$ |
| $3.00\times10^6$ | $9.28\times10^4$ | $9.28\times10^3$ | $8.62\times10^3$ | $6.64\times10^5$ | $2.12\times10^1$ | $1.46\times10^{-16}$ | $1.06\times10^{-1}$ |
| $3.00\times10^6$ | $1.03\times10^5$ | $1.03\times10^4$ | $5.68\times10^3$ | $1.53\times10^6$ | 3.99 | $1.46\times10^{-16}$ | $1.06\times10^{-1}$ |
| $3.00\times10^6$ | $1.15\times10^5$ | $1.15\times10^4$ | $3.59\times10^3$ | $3.82\times10^6$ | $6.39\times10^{-1}$ | $1.46\times10^{-16}$ | $1.06\times10^{-1}$ |
| $3.00\times10^6$ | $1.27\times10^5$ | $1.27\times10^4$ | $2.16\times10^3$ | $1.06\times10^7$ | $8.40\times10^{-2}$ | $1.46\times10^{-16}$ | $1.06\times10^{-1}$ |
| $3.00\times10^6$ | $1.41\times10^5$ | $1.41\times10^4$ | $1.21\times10^3$ | $3.35\times10^7$ | $8.35\times10^{-3}$ | $1.46\times10^{-16}$ | $1.06\times10^{-1}$ |
| $3.00\times10^6$ | $1.57\times10^5$ | $1.57\times10^4$ | $6.49\times10^2$ | $1.17\times10^8$ | $6.80\times10^{-4}$ | $1.46\times10^{-16}$ | $1.06\times10^{-1}$ |
| $3.00\times10^6$ | $1.75\times10^5$ | $1.75\times10^4$ | $3.20\times10^2$ | $4.82\times10^8$ | $4.03\times10^{-5}$ | $1.46\times10^{-16}$ | $1.06\times10^{-1}$ |
| $3.00\times10^6$ | $1.94\times10^5$ | $1.94\times10^4$ | $1.47\times10^2$ | $2.27\times10^9$ | $1.81\times10^{-6}$ | $1.46\times10^{-16}$ | $1.06\times10^{-1}$ |
| $3.00\times10^6$ | $2.16\times10^5$ | $2.16\times10^4$ | $6.13\times10^1$ | $1.31\times10^{10}$ | $5.44\times10^{-8}$ | $1.46\times10^{-16}$ | $1.06\times10^{-1}$ |

**Table S3: Representative results for $B_o = 2.5 \times 10^{12}$ T.**

| MQN Mass (kg) | Altitude $h$ (m) | $\Delta h$ (m) | Frequency (Hz) | $\tau\_down$ (s) | RF power (W) | Flux ($m^{-2}y^{-1}sr^{-1}$) | Number/y for all directions |
|---|---|---|---|---|---|---|---|
| $3.00 \times 10^6$ | $6.14 \times 10^3$ | $6.14 \times 10^2$ | $2.75 \times 10^5$ | $4.17 \times 10^2$ | $3.44 \times 10^7$ | $7.27 \times 10^{-19}$ | $5.29 \times 10^{-4}$ |
| $3.00 \times 10^6$ | $7.14 \times 10^3$ | $7.14 \times 10^2$ | $2.64 \times 10^5$ | $4.52 \times 10^2$ | $2.93 \times 10^7$ | $7.27 \times 10^{-19}$ | $5.29 \times 10^{-4}$ |
| $3.00 \times 10^6$ | $8.14 \times 10^3$ | $8.14 \times 10^2$ | $2.54 \times 10^5$ | $4.89 \times 10^2$ | $2.50 \times 10^7$ | $7.27 \times 10^{-19}$ | $5.29 \times 10^{-4}$ |
| $3.00 \times 10^6$ | $9.14 \times 10^3$ | $9.14 \times 10^2$ | $2.44 \times 10^5$ | $5.30 \times 10^2$ | $2.13 \times 10^7$ | $7.27 \times 10^{-19}$ | $5.29 \times 10^{-4}$ |
| $3.00 \times 10^6$ | $1.02 \times 10^4$ | $1.02 \times 10^3$ | $2.35 \times 10^5$ | $5.74 \times 10^2$ | $1.82 \times 10^7$ | $7.27 \times 10^{-19}$ | $5.29 \times 10^{-4}$ |
| $3.00 \times 10^6$ | $1.13 \times 10^4$ | $1.13 \times 10^3$ | $2.23 \times 10^5$ | $6.34 \times 10^2$ | $1.49 \times 10^7$ | $7.27 \times 10^{-19}$ | $5.29 \times 10^{-4}$ |
| $3.00 \times 10^6$ | $1.25 \times 10^4$ | $1.25 \times 10^3$ | $2.12 \times 10^5$ | $7.00 \times 10^2$ | $1.22 \times 10^7$ | $7.27 \times 10^{-19}$ | $5.29 \times 10^{-4}$ |
| $3.00 \times 10^6$ | $1.39 \times 10^4$ | $1.39 \times 10^3$ | $2.02 \times 10^5$ | $7.73 \times 10^2$ | $1.00 \times 10^7$ | $7.27 \times 10^{-19}$ | $5.29 \times 10^{-4}$ |
| $3.00 \times 10^6$ | $1.55 \times 10^4$ | $1.55 \times 10^3$ | $1.90 \times 10^5$ | $8.72 \times 10^2$ | $7.88 \times 10^6$ | $7.27 \times 10^{-19}$ | $5.29 \times 10^{-4}$ |
| $3.00 \times 10^6$ | $1.72 \times 10^4$ | $1.72 \times 10^3$ | $1.77 \times 10^5$ | $1.00 \times 10^3$ | $5.96 \times 10^6$ | $7.27 \times 10^{-19}$ | $5.29 \times 10^{-4}$ |
| $3.00 \times 10^6$ | $1.91 \times 10^4$ | $1.91 \times 10^3$ | $1.64 \times 10^5$ | $1.17 \times 10^3$ | $4.34 \times 10^6$ | $7.27 \times 10^{-19}$ | $5.29 \times 10^{-4}$ |
| $3.00 \times 10^6$ | $2.12 \times 10^4$ | $2.12 \times 10^3$ | $1.51 \times 10^5$ | $1.38 \times 10^3$ | $3.15 \times 10^6$ | $7.27 \times 10^{-19}$ | $5.29 \times 10^{-4}$ |
| $3.00 \times 10^6$ | $2.36 \times 10^4$ | $2.36 \times 10^3$ | $1.37 \times 10^5$ | $1.68 \times 10^3$ | $2.12 \times 10^6$ | $7.27 \times 10^{-19}$ | $5.29 \times 10^{-4}$ |
| $3.00 \times 10^6$ | $2.62 \times 10^4$ | $2.62 \times 10^3$ | $1.23 \times 10^5$ | $2.09 \times 10^3$ | $1.37 \times 10^6$ | $7.27 \times 10^{-19}$ | $5.29 \times 10^{-4}$ |
| $3.00 \times 10^6$ | $2.91 \times 10^4$ | $2.91 \times 10^3$ | $1.10 \times 10^5$ | $2.60 \times 10^3$ | $8.83 \times 10^5$ | $7.27 \times 10^{-19}$ | $5.29 \times 10^{-4}$ |
| $3.00 \times 10^6$ | $3.24 \times 10^4$ | $3.24 \times 10^3$ | $9.67 \times 10^4$ | $3.37 \times 10^3$ | $5.26 \times 10^5$ | $7.27 \times 10^{-19}$ | $5.29 \times 10^{-4}$ |
| $3.00 \times 10^6$ | $3.60 \times 10^4$ | $3.60 \times 10^3$ | $8.33 \times 10^4$ | $4.55 \times 10^3$ | $2.90 \times 10^5$ | $7.27 \times 10^{-19}$ | $5.29 \times 10^{-4}$ |
| $3.00 \times 10^6$ | $4.00 \times 10^4$ | $4.00 \times 10^3$ | $7.11 \times 10^4$ | $6.25 \times 10^3$ | $1.53 \times 10^5$ | $7.27 \times 10^{-19}$ | $5.29 \times 10^{-4}$ |
| $3.00 \times 10^6$ | $4.44 \times 10^4$ | $4.44 \times 10^3$ | $5.94 \times 10^4$ | $8.94 \times 10^3$ | $7.48 \times 10^4$ | $7.27 \times 10^{-19}$ | $5.29 \times 10^{-4}$ |
| $3.00 \times 10^6$ | $4.93 \times 10^4$ | $4.93 \times 10^3$ | $4.87 \times 10^4$ | $1.33 \times 10^4$ | $3.38 \times 10^4$ | $7.27 \times 10^{-19}$ | $5.29 \times 10^{-4}$ |
| $3.00 \times 10^6$ | $5.48 \times 10^4$ | $5.48 \times 10^3$ | $3.91 \times 10^4$ | $2.06 \times 10^4$ | $1.41 \times 10^4$ | $7.27 \times 10^{-19}$ | $5.29 \times 10^{-4}$ |
| $3.00 \times 10^6$ | $6.09 \times 10^4$ | $6.09 \times 10^3$ | $3.05 \times 10^4$ | $3.39 \times 10^4$ | $5.20 \times 10^3$ | $7.27 \times 10^{-19}$ | $5.29 \times 10^{-4}$ |
| $3.00 \times 10^6$ | $6.77 \times 10^4$ | $6.77 \times 10^3$ | $2.33 \times 10^4$ | $5.81 \times 10^4$ | $1.77 \times 10^3$ | $7.27 \times 10^{-19}$ | $5.29 \times 10^{-4}$ |
| $3.00 \times 10^6$ | $7.52 \times 10^4$ | $7.52 \times 10^3$ | $1.73 \times 10^4$ | $1.05 \times 10^5$ | $5.38 \times 10^2$ | $7.27 \times 10^{-19}$ | $5.29 \times 10^{-4}$ |
| $3.00 \times 10^6$ | $8.36 \times 10^4$ | $8.36 \times 10^3$ | $1.23 \times 10^4$ | $2.08 \times 10^5$ | $1.39 \times 10^2$ | $7.27 \times 10^{-19}$ | $5.29 \times 10^{-4}$ |
| $3.00 \times 10^6$ | $9.28 \times 10^4$ | $9.28 \times 10^3$ | $8.45 \times 10^3$ | $4.42 \times 10^5$ | $3.06 \times 10^1$ | $7.27 \times 10^{-19}$ | $5.29 \times 10^{-4}$ |
| $3.00 \times 10^6$ | $1.03 \times 10^5$ | $1.03 \times 10^4$ | $5.56 \times 10^3$ | $1.02 \times 10^6$ | $5.76$ | $7.27 \times 10^{-19}$ | $5.29 \times 10^{-4}$ |
| $3.00 \times 10^6$ | $1.15 \times 10^5$ | $1.15 \times 10^4$ | $3.52 \times 10^3$ | $2.55 \times 10^6$ | $9.22 \times 10^{-1}$ | $7.27 \times 10^{-19}$ | $5.29 \times 10^{-4}$ |
| $3.00 \times 10^6$ | $1.27 \times 10^5$ | $1.27 \times 10^4$ | $2.12 \times 10^3$ | $7.03 \times 10^6$ | $1.21 \times 10^{-1}$ | $7.27 \times 10^{-19}$ | $5.29 \times 10^{-4}$ |
| $3.00 \times 10^6$ | $1.41 \times 10^5$ | $1.41 \times 10^4$ | $1.19 \times 10^3$ | $2.23 \times 10^7$ | $1.20 \times 10^{-2}$ | $7.27 \times 10^{-19}$ | $5.29 \times 10^{-4}$ |
| $3.00 \times 10^6$ | $1.57 \times 10^5$ | $1.57 \times 10^4$ | $6.36 \times 10^2$ | $7.81 \times 10^7$ | $9.81 \times 10^{-4}$ | $7.27 \times 10^{-19}$ | $5.29 \times 10^{-4}$ |
| $3.00 \times 10^6$ | $1.75 \times 10^5$ | $1.75 \times 10^4$ | $3.14 \times 10^2$ | $3.21 \times 10^8$ | $5.82 \times 10^{-5}$ | $7.27 \times 10^{-19}$ | $5.29 \times 10^{-4}$ |
| $3.00 \times 10^6$ | $1.94 \times 10^5$ | $1.94 \times 10^4$ | $1.44 \times 10^2$ | $1.51 \times 10^9$ | $2.61 \times 10^{-6}$ | $7.27 \times 10^{-19}$ | $5.29 \times 10^{-4}$ |
| $3.00 \times 10^6$ | $2.16 \times 10^5$ | $2.16 \times 10^4$ | $6.07 \times 10^1$ | $8.56 \times 10^9$ | $8.17 \times 10^{-8}$ | $7.27 \times 10^{-19}$ | $5.29 \times 10^{-4}$ |

**Table S4: Representative results for $B_o = 3.0 \times 10^{12}$ T.**

| MQN Mass (kg) | Altitude $h$ (m) | $\Delta h$ (m) | Frequency (Hz) | $\tau\_\text{down}$ (s) | RF power (W) | Flux ($m^{-2}y^{-1}sr^{-1}$) | Number/y for all directions |
|---|---|---|---|---|---|---|---|
| 3.00 | $1.41\times10^5$ | $1.41\times10^4$ | $1.18\times10^5$ | $1.57\times10^5$ | $1.68\times10^{-6}$ | $9.55\times10^{-18}$ | $6.60\times10^{-5}$ |
| 3.00 | $1.27\times10^5$ | $1.27\times10^4$ | $2.08\times10^5$ | $5.06\times10^4$ | $1.62\times10^{-5}$ | $9.55\times10^{-18}$ | $6.57\times10^{-5}$ |
| 3.00 | $1.15\times10^5$ | $1.15\times10^4$ | $3.46\times10^5$ | $1.83\times10^4$ | $1.24\times10^{-4}$ | $9.55\times10^{-18}$ | $5.89\times10^{-5}$ |
| 3.00 | $1.03\times10^5$ | $1.03\times10^4$ | $5.52\times10^5$ | $7.20\times10^3$ | $8.02\times10^{-4}$ | $9.55\times10^{-18}$ | $5.28\times10^{-5}$ |
| $3.00\times10^{-3}$ | $7.52\times10^4$ | $7.52\times10^3$ | $1.68\times10^7$ | $7.81\times10^1$ | $6.81\times10^{-4}$ | $1.80\times10^{-17}$ | $5.06\times10^{-5}$ |
| 3.00 | $9.28\times10^4$ | $9.28\times10^3$ | $8.30\times10^5$ | $3.18\times10^3$ | $4.10\times10^{-3}$ | $9.55\times10^{-18}$ | $4.74\times10^{-5}$ |
| $3.00\times10^{-3}$ | $6.77\times10^4$ | $6.77\times10^3$ | $2.24\times10^7$ | $4.39\times10^1$ | $2.16\times10^{-3}$ | $1.80\times10^{-17}$ | $4.69\times10^{-5}$ |
| $3.00\times10^{-1}$ | $9.28\times10^4$ | $9.28\times10^3$ | $1.80\times10^6$ | $1.45\times10^3$ | $9.15\times10^{-4}$ | $9.69\times10^{-18}$ | $4.31\times10^{-5}$ |
| 3.00 | $8.36\times10^4$ | $8.36\times10^3$ | $1.21\times10^6$ | $1.49\times10^3$ | $1.86\times10^{-2}$ | $9.55\times10^{-18}$ | $4.25\times10^{-5}$ |
| $3.00\times10^{-3}$ | $6.09\times10^4$ | $6.09\times10^3$ | $2.87\times10^7$ | $2.67\times10^1$ | $5.84\times10^{-3}$ | $1.80\times10^{-17}$ | $4.21\times10^{-5}$ |
| $3.00\times10^3$ | $2.16\times10^5$ | $2.16\times10^4$ | $5.99\times10^2$ | $6.11\times10^8$ | $1.11\times10^{-9}$ | $2.39\times10^{-18}$ | $4.03\times10^{-5}$ |
| $3.00\times10^{-1}$ | $8.36\times10^4$ | $8.36\times10^3$ | $2.61\times10^6$ | $6.95\times10^2$ | $3.99\times10^{-3}$ | $9.69\times10^{-18}$ | $3.87\times10^{-5}$ |
| 3.00 | $7.52\times10^4$ | $7.52\times10^3$ | $1.70\times10^6$ | $7.60\times10^2$ | $7.20\times10^{-2}$ | $9.55\times10^{-18}$ | $3.82\times10^{-5}$ |
| $3.00\times10^{-3}$ | $5.48\times10^4$ | $5.48\times10^3$ | $3.50\times10^7$ | $1.79\times10^1$ | $1.30\times10^{-2}$ | $1.80\times10^{-17}$ | $3.78\times10^{-5}$ |
| $3.00\times10^{-2}$ | $8.36\times10^4$ | $8.36\times10^3$ | $5.61\times10^6$ | $3.24\times10^2$ | $8.55\times10^{-4}$ | $1.01\times10^{-17}$ | $3.62\times10^{-5}$ |
| $3.00\times10^3$ | $1.94\times10^5$ | $1.94\times10^4$ | $1.42\times10^3$ | $1.08\times10^8$ | $3.55\times10^{-8}$ | $2.39\times10^{-18}$ | $3.59\times10^{-5}$ |
| $3.00\times10^{-1}$ | $7.52\times10^4$ | $7.52\times10^3$ | $3.65\times10^6$ | $3.53\times10^2$ | $1.54\times10^{-2}$ | $9.69\times10^{-18}$ | $3.47\times10^{-5}$ |
| $3.00\times10^{-4}$ | $6.09\times10^4$ | $6.09\times10^3$ | $5.75\times10^7$ | $1.43\times10^1$ | $9.48\times10^{-4}$ | $1.49\times10^{-17}$ | $3.47\times10^{-5}$ |
| 3.00 | $6.77\times10^4$ | $6.77\times10^3$ | $2.29\times10^6$ | $4.18\times10^2$ | $2.38\times10^{-1}$ | $9.55\times10^{-18}$ | $3.42\times10^{-5}$ |
| $3.00\times10^{-3}$ | $4.93\times10^4$ | $4.93\times10^3$ | $4.10\times10^7$ | $1.30\times10^1$ | $2.45\times10^{-2}$ | $1.80\times10^{-17}$ | $3.40\times10^{-5}$ |
| $3.00\times10^{-2}$ | $7.52\times10^4$ | $7.52\times10^3$ | $7.79\times10^6$ | $1.68\times10^2$ | $3.18\times10^{-3}$ | $1.01\times10^{-17}$ | $3.25\times10^{-5}$ |
| $3.00\times10^3$ | $1.75\times10^5$ | $1.75\times10^4$ | $3.09\times10^3$ | $2.29\times10^7$ | $7.92\times10^{-7}$ | $2.39\times10^{-18}$ | $3.21\times10^{-5}$ |
| $3.00\times10^{-1}$ | $6.77\times10^4$ | $6.77\times10^3$ | $4.93\times10^6$ | $1.95\times10^2$ | $5.10\times10^{-2}$ | $9.69\times10^{-18}$ | $3.12\times10^{-5}$ |
| $3.00\times10^{-4}$ | $5.48\times10^4$ | $5.48\times10^3$ | $6.75\times10^7$ | $1.04\times10^1$ | $1.79\times10^{-3}$ | $1.49\times10^{-17}$ | $3.12\times10^{-5}$ |
| 3.00 | $6.09\times10^4$ | $6.09\times10^3$ | $3.00\times10^6$ | $2.44\times10^2$ | $6.96\times10^{-1}$ | $9.55\times10^{-18}$ | $3.08\times10^{-5}$ |
| $3.00\times10^{-2}$ | $6.77\times10^4$ | $6.77\times10^3$ | $1.05\times10^7$ | $9.24\times10^1$ | $1.05\times10^{-2}$ | $1.01\times10^{-17}$ | $2.92\times10^{-5}$ |
| $3.00\times10^3$ | $1.57\times10^5$ | $1.57\times10^4$ | $6.27\times10^3$ | $5.58\times10^6$ | $1.34\times10^{-5}$ | $2.39\times10^{-18}$ | $2.87\times10^{-5}$ |
| $3.00\times10^{-1}$ | $6.09\times10^4$ | $6.09\times10^3$ | $6.44\times10^6$ | $1.14\times10^2$ | $1.49\times10^{-1}$ | $9.69\times10^{-18}$ | $2.80\times10^{-5}$ |
| $3.00\times10^1$ | $1.57\times10^5$ | $1.57\times10^4$ | $2.90\times10^4$ | $1.21\times10^6$ | $6.14\times10^{-7}$ | $2.92\times10^{-18}$ | $2.80\times10^{-5}$ |
| 3.00 | $5.48\times10^4$ | $5.48\times10^3$ | $3.80\times10^6$ | $1.52\times10^2$ | 1.81 | $9.55\times10^{-18}$ | $2.76\times10^{-5}$ |
| $3.00\times10^{-2}$ | $6.09\times10^4$ | $6.09\times10^3$ | $1.36\times10^7$ | $5.51\times10^1$ | $2.95\times10^{-2}$ | $1.01\times10^{-17}$ | $2.62\times10^{-5}$ |
| $3.00\times10^3$ | $1.41\times10^5$ | $1.41\times10^4$ | $1.17\times10^4$ | $1.59\times10^6$ | $1.64\times10^{-4}$ | $2.39\times10^{-18}$ | $2.56\times10^{-5}$ |
| $3.00\times10^{-1}$ | $5.48\times10^4$ | $5.48\times10^3$ | $8.10\times10^6$ | $7.19\times10^1$ | $3.73\times10^{-1}$ | $9.69\times10^{-18}$ | $2.52\times10^{-5}$ |
| $3.00\times10^1$ | $1.41\times10^5$ | $1.41\times10^4$ | $5.49\times10^4$ | $3.38\times10^5$ | $7.84\times10^{-6}$ | $2.92\times10^{-18}$ | $2.50\times10^{-5}$ |